\documentclass[12pt]{article}
\usepackage{amsmath,amssymb,mathrsfs,framed}
\usepackage[colorlinks]{hyperref}
\usepackage{color}

\usepackage[all]{xy}

\newcommand{\bnu}{{\boldsymbol{\nu}}}

\newcommand{\bgamma}{{\boldsymbol{\gamma}}}

\newcommand{\bn}{{\mathbf{n}}}
\newcommand{\bx}{{\mathbf{x}}}

\newcommand{\de}{{\mathrm{d}}}

\newcommand{\rem}[1]{}

\makeatletter

\@addtoreset{figure}{section}
\def\thefigure{\thesection.\@arabic\c@figure}
\def\fps@figure{h, t}
\@addtoreset{table}{bsection}
\def\thetable{\thesection.\@arabic\c@table}
\def\fps@table{h, t}
\@addtoreset{equation}{section}

\makeatother

\newcommand{\todo}[1]{\vspace{5 mm}\par \noindent
\framebox{\begin{minipage}[c]{0.95 \textwidth}
\tt #1 \end{minipage}}\vspace{5 mm}\par}
\begin{document}

\newtheorem{theorem}{Theorem}[section]
\newtheorem{definition}[theorem]{Definition}
\newtheorem{lemma}[theorem]{Lemma}
\newtheorem{remark}[theorem]{Remark}
\newtheorem{proposition}[theorem]{Proposition}
\newtheorem{corollary}[theorem]{Corollary}
\newtheorem{example}[theorem]{Example}

\def\below#1#2{\mathrel{\mathop{#1}\limits_{#2}}}

%%%%%%%%%%%%%%%%%%%%%%%%%%%%%%%%%%%%%%%%%%%%%%%%%%%%%%%%%%%%%%%%%%%%%%%%%%%%%%%
%%%%%%%%

%%%%%%%%%%%%%%%%%%%%%%%%%%%%%%%%%%%%%%%%%%%%%%%%%%%%%%%%%%%%%%%%%%%%%%%%%%%%%%%
%%%%%%%%

\title{\vspace{-.4cm}Euler-Poincar\'e approaches to nematodynamics}
\author{Fran\c{c}ois Gay-Balmaz$^{1}$, Tudor S. Ratiu$^{2}$, Cesare Tronci$^{2,3}$}
\addtocounter{footnote}{1} 
\footnotetext{\scriptsize Laboratoire de 
M\'et\'eorologie Dynamique, \'Ecole Normale Sup\'erieure/CNRS, Paris, France. 
\texttt{gaybalma@lmd.ens.fr}
\addtocounter{footnote}{1} }
\footnotetext{\scriptsize Section de
Math\'ematiques, \'Ecole Polytechnique F\'ed\'erale de
Lausanne. CH--1015 Lausanne. Switzerland. TSR was partially supported by Swiss NSF grant 200020-126630 and by the government grant of the Russian Federation for support of research projects implemented by leading scientists, Lomonosov Moscow State University under the agreement No. 11.G34.31.0054. 
\texttt{tudor.ratiu@epfl.ch}, 
\addtocounter{footnote}{1} }
\footnotetext{\scriptsize Department of Mathematics, University of Surrey. Guildford GU2 7XH. United Kingdom. \texttt{c.tronci@surrey.ac.uk}
\addtocounter{footnote}{1} }

\date{ }
\maketitle

\makeatother
%\begin{center} DRAFT \end{center}
\maketitle

%|||-------------------text width----------------------|||

%\noindent \textbf{AMS Classification:}

%\noindent \textbf{Keywords:}

\begin{abstract}
Nematodynamics is the orientation dynamics of flowless liquid-crystals. We show how Euler-Poincar\'e reduction produces a unifying framework for various theories, including Ericksen-Leslie, 
Luhiller-Rey, and Eringen's micropolar theory. In particular, we show that these theories are all compatible with each other and some of them allow for more general configurations involving a non vanishing disclination density. All results are also extended to flowing liquid crystals.
\end{abstract}

{\footnotesize
\tableofcontents
}

%-------------------------------------------------------
%--------------------------------------------------------

\bigskip

\section{Introduction}

\subsection{Ericksen-Leslie theory vs. Eringen's micropolar theory}
A flowless nematic liquid crystal possesses orientational dynamics that involves the director variable $\bn(\bx,t)$ along with its associated angular velocity. {\color{black} The director $\mathbf{n}$ is a unit vector, i.e., $\left|\mathbf{n}\right|=1$, that is unsigned, so that $\mathbf{n}$ is identified
with $- \mathbf{n}$ (the head is identified with the tail).} This orientational dynamics is often called nematodynamics and its equations are subject of continuing research. Depending on whether inertial effects are considered or not, different theories are obtained. When these effects are neglected, dissipation plays a fundamental role and the problem is often approached from a kinetic theory point of view \cite{DoEd1988}. Although most of realistic situations involve negligible inertia, nematodynamics theories are best understood when inertial effects are taken into account. For the purposes of this paper, we shall consider the ideal cases when dissipation is absent and inertial effects are not negligible. For example, in the ideal case of zero dissipation, the celebrated Ericksen-Leslie equation
\begin{equation*}
J\frac{d^2\mathbf{n} }{dt^2}-\left(\mathbf{n\cdot h}+J\,\mathbf{n}\cdot\frac{d^2\mathbf{n} }{dt^2}\right)\mathbf{n}+\mathbf{h}=0
\end{equation*}
 has been given a well defined Hamiltonian structure; see
 \cite{Volovick1980}. In the above equation, $J$ is the \emph{microinertia constant} and the \emph{molecular field} 
\[
\mathbf{h}:=\frac{\delta F}{\delta \bn}-\frac{\partial}{\partial x^i}\frac{\delta F}{\delta (\partial_{x^i}\bn)}
\]
is expressed in terms of the Frank $F$ energy, which in turn is given by
 \begin{align*}
F(\mathbf{n},{\nabla}\mathbf{n})&:=K_2\underbrace{(\mathbf{n}\cdot\operatorname{curl}\mathbf{n})}_{\textbf{chirality}}+\frac{1}{2}K_{11}\underbrace{(\operatorname{div}\mathbf{n})^2}_{\textbf{splay}}+\frac{1}{2}K_{22}\underbrace{(\mathbf{n}\cdot\operatorname{curl}\mathbf{n})^2}_{\textbf{twist}}
+\frac{1}{2}K_{33}\underbrace{\|\mathbf{n}\times\operatorname{curl}\mathbf{n}\|^2}_{\textbf{bend}},
\end{align*}
where each term possesses a precise physical meaning, as indicated above.

The Ericksen-Leslie equation for the dynamics of nematic liquid crystals is a widely accepted model since it has been experimentally validated through various measurements 
\cite{deGennes1971,Chandra1992,ReyDenn}. However, orientational defects (\emph{disclinations}), make this model unreliable. For
example, if
defects are present (singularities in the orientational order), uniaxiality may be lost and the model becomes biaxial
thereby eliminating the usefulness of the director field 
$\mathbf{n}$: one needs some other order parameter.

There are other approaches that take into account defect dynamics. 
The most general one was proposed by Eringen \cite{Eringen1997} 
in the context of microfluid motion that includes liquid crystals. 
The Eringen model admits molecular shape changes by introducing
a microinertia tensor $j$, whose dynamics is coupled to 
the \emph{wryness tensor} $\gamma$, which is often expressed in terms of $(\nabla\mathbf{n})\times\mathbf{n}$ when disclinations are absent 
\cite{Eringen1997}. {\color{black} In this paper, the notation $\nabla  \mathbf{n} $ stands for the vector valued one-form that is given by the derivative of the director $\mathbf{n} $, viewed as a map with values in $ S ^2 \subset \mathbb{R}  ^3$.}

Nematic liquid crystals are well known to be a typical example of 
microfluids. Several unsuccessful attempts have been made to show 
how the Ericksen-Leslie (EL) description arises from Eringen's 
micropolar theory. The most notable one is the relation 
$\boldsymbol{\gamma}=(\nabla\mathbf{n})\times\mathbf{n}$ proposed 
by Eringen \cite[formula (11.2)]{Eringen1997}; it does not
yield the EL equations \cite{Le1979} as shown in
\cite[Theorem 8.11]{GBRa2009} by two different methods (symmetry considerations and a direct computation). No good formula exists 
that expresses $\boldsymbol{\gamma}$ in terms of the director
$\mathbf{n}$. This difficulty arises even in the simplest ideal case when dissipation is totally absent.

\subsection{The gauge-theory approach}
\noindent
Reduction theory has recently contributed to the understanding of
defect dynamics \cite{Ho2002,GBRa2009}; it leads directly to the gauge-theory approach  \cite{Volovick1980} and applies to very general systems since it incorporates defect dynamics in various
models, such as frustrated spin glasses 
\cite{HoKu1988,Volovick1980}, for example. In this gauge theoretical 
setting, one interprets the wryness tensor $\boldsymbol{\gamma}$ as 
the magnetic potential of a Yang-Mills field (i.e., a 
\emph{connection one-form}) taking values in the Lie algebra 
$\mathfrak{so}(3)$ of antisymmetric $3 \times 3$ matrices (usually 
identified with vectors in $\mathbb{R}^3$) of the proper rotation 
group $SO(3)$. The one-form $\boldsymbol\gamma$ is also known as 
`spatial rotational strain' \cite{Ho2002};  it usually expresses the amount 
by which a specified director field rotates under an infinitesimal 
displacement. The gauge potential $\boldsymbol{\gamma}$ can be 
written in terms of a chosen basis $\{\mathbf{e}_a\mid a=1,2,3\}$ 
of $\mathbb{R}^3 \simeq\mathfrak{so}(3)$ as
\[
\bgamma=\bgamma_i\,\de x^i=\gamma_i^a\,\mathbf{e}_a\,\de x^i.
\]
Then, its corresponding magnetic vector field (the \textit{curvature} of $\boldsymbol{\gamma}$) is given componentwise by
\begin{equation}\label{Magnetic-Field}
\boldsymbol{B}^i=\epsilon^{ijk}\!\left(\partial_j\bgamma_k+\bgamma_j\times\bgamma_k\right);
\end{equation}
summation over repeated indices is understood and two-forms on 
physical space $\mathbb{R}^3$ have been identified with vector 
fields. The absence of disclinations in the gauge theory approach
(see \cite{Volovick1980}) is characterized by the vanishing of
the magnetic field $\boldsymbol{B}$ and not by the vanishing
of the vector potential $\boldsymbol{\gamma}$. \textcolor{black}{Therefore, if
$\boldsymbol{\gamma} \neq 0$ must be compatible with EL dynamics, then one requires
 that $\boldsymbol{B}=0$. In \cite{Volovick1980} it
was shown that $\boldsymbol{B}=0$ comes down to the homogeneous 
initial condition $\boldsymbol{B}_0=0$ (that is, for example, $\boldsymbol{\gamma}_0=0$)}. If 
$\boldsymbol{B}\neq 0$, then the gauge theoretical model
confers the director formulation the possibility of including non-trivial disclinations.

Although Eringen's micropolar theory appears  to possess a 
gauge-theoretical 
formulation due to the gauge invariant relation $\nabla\bn=\bn\times\bgamma$, Eringen's final choice of identifying the wryness tensor $\bgamma$ with $(\nabla\mathbf{n})\times\mathbf{n}$ leads to problems arising from the fact that  the latter expression does not transform as a magnetic potential under 
gauge transformations (i.e., it is not gauge-invariant; see  \cite[Lemma 8.10]{GBRa2009}). Still, Eringen's theory exhibits many properties analogous to gauge-theoretical models. In addition, the simultaneous presence of the
microinertia and wryness tensors in Eringen's model accounts for
the interaction of molecular shape with a non-vanishing disclination density $\boldsymbol{B}$.

\subsection{The role of reduction by symmetry}
The comments above are the main motivation for the present work.
We shall employ Euler-Poincar\'e variational methods to produce a framework that incorporates defect dynamics in continuum systems with broken internal symmetry (e.g., liquid crystals) and show
that Eringen's micropolar theory contains Ericksen-Leslie dynamics. 
The main idea is that the gradient of the identity
\[
\mathbf{n}(\mathbf{x},t)=\chi(\mathbf{x},t)\,\mathbf{e}_3,
\]
relating director dynamics to the dynamics of the rotation matrix 
$\chi(\mathbf{x},t)\in SO(3)$ in EL theory, is
\[
\nabla\mathbf{n}=(\nabla\chi)\mathbf{e}_3
=(\nabla\chi)\chi^{-1\,}\mathbf{n}\,,
\]
where $\mathbf{e}_3:=(0,0,1)$, the unit vector on the positive
vertical axis. Next, note that the new variable 
\begin{equation}
\widehat{\bgamma}=-(\nabla\chi)\chi^{-1}
\label{bgamma}
\end{equation}
is a $\mathfrak{so}(3)$-valued connection one-form  
\cite{Ho2002,GBRa2009}.
\begin{remark}[The hat map]\normalfont
In this paper we shall often make use of the following `hat map' 
isomorphism of Lie algebras $\;\widehat\;:\Bbb{R}^3\to\mathfrak{so}(3)$ (see e.g., \cite{MaRa2004}) 
\[
\widehat\nu_{jk}=-\epsilon_{jki\,}\nu_i\,,\qquad \nu_i=-\frac12\epsilon_{ijk\,}\widehat\nu_{jk}
\,,
\]
holding for any vector $\bnu\in\Bbb{R}^3$ and its associated antisymmetric matrix $\widehat{\boldsymbol{\nu}}\in\mathfrak{so}(3)$, so that $\widehat{\boldsymbol{\nu}}\mathbf{u}=\bnu\times\mathbf{u}$ for any vector $\mathbf{u}\in\Bbb{R}^3$. In the formulas above, summation of repeated indexes is used. {\color{black} In particular,  applying the inverse of the hat map to the relation \eqref{bgamma} yields Eringen's wryness tensor $\bgamma_i\,\de x^i$ \cite{Eringen1997} (see discussions below).}
\end{remark}
The
two relations above yield hence 
$\nabla\mathbf{n}=\mathbf{n}\times\boldsymbol{\gamma}$ which, when
used in the EL equations introduce $\boldsymbol{\gamma}$ as a new
dynamical variable. Note that the choice $\boldsymbol{\gamma}=
(\nabla\mathbf{n})\times\mathbf{n}$ satisfies the relation
$\nabla\mathbf{n}=\mathbf{n}\times\boldsymbol{\gamma}$; however
this latter relation determines $\boldsymbol{\gamma}$ only up
to a component parallel to $\mathbf{n}$. Although this was noticed by Eringen in \cite{Eringen1993}, he did not observe that, as a consequence, 
$\boldsymbol{\gamma}$ is not a function of $\mathbf{n}$ alone; in
fact all three columns of the matrix $\chi(\mathbf{x},t)$ are needed
to specify $\boldsymbol{\gamma}$.

The second main observation is that a different symmetry reduction
of the same material Lagrangian yields a new set of
equations for nematodynamics. This is due to a symmetry that seems
not to have been exploited before. We will show that these new 
reduced equations are completely equivalent to the Ericksen-Leslie
equations. However, this new system permits the
description of disclination density dynamics, something that the Ericksen-Leslie equations could not handle explicitly.

All the above considerations hold independent of the background 
fluid motion (the \textit{macromotion}); they affect only the (orientational) 
micromotion. Because of this, we will treat for a while only
liquid crystals with zero background fluid motion since this allows
us to concentrate on the main points of the relation between the
Ericksen-Leslie and Eringen models. At the end of the paper we shall
quickly present liquid crystals with macromotion.

\section{Euler-Poincar\'e reduction for nematic systems}

Nematic systems are continuum media in which each particle of the system carries an orientation, that is, a (time-dependent) rotational matrix $\chi(\mathbf{x},t)\in SO(3)$ is attached at each point $\bx$ in physical space $\mathbb{R}^3$. {\color{black} More specifically, we shall consider a medium occupying a bounded domain $\mathcal{D}\subset\Bbb{R}^3$, so that $\bx\in\mathcal{D}$ and, at each time $t$, $\chi$ is a map from $\mathcal{D}$ to $SO(3)$; the set of all such maps will be denoted by $\mathcal{F(D},SO(3))$.} When dissipation is neglected, a nematic system can always be described by the Euler-Lagrange equations
\[
\frac{\de}{\de t}\frac{\delta \mathcal{L}}{\delta \dot\chi}-\frac{\delta \mathcal{L}}{\delta \chi}=0
\]
(the notation $\dot\chi$ stands for the partial time derivative 
$\partial_t\chi$) arising from Hamilton's principle
\[
\delta\int_{t_1}^{t_2}\mathcal{L} ( \chi , \dot \chi )\,\de t=0
\,,
\]
for any variation $\delta\chi(t) \in \mathcal{F}(\mathcal{D},SO(3))$
satisfying $\delta\chi(t_1) = \delta\chi(t_2) = 0$,
where the Lagrangian functional $\mathcal{L}$ involves a Lagrangian density $\mathscr{L}$ as follows:
\[
\mathcal{L} ( \chi , \dot \chi )=\int_\mathcal{D}\mathscr{L}(\chi(\bx,t) , \dot \chi(\bx,t))\,\de^3\mathbf{x}
\,.
\]
As usual, we think of $\chi$ as a 
$\mathcal{F}(\mathcal{D}, SO(3))$-valued curve, where 
$\mathcal{D} \subset \mathbb{R}^3$ is the physical domain of the 
liquid crystal and $\mathcal{F}(\mathcal{D}, SO(3))$ denotes the 
group (relative to point-wise multiplication) of smooth 
$SO(3)$-valued functions on $\mathcal{D}$. Thus the Lagrangian 
$\mathcal{L}$ is defined on the state space 
$T \mathcal{F}(\mathcal{D}, SO(3))$
and we denote a tangent vector to $\mathcal{F}(\mathcal{D}, SO(3))$ 
at $\chi$ by $(\chi, \dot{\chi}) \in T_\chi 
\mathcal{F}(\mathcal{D}, SO(3))$.

\subsection{Continuum systems with rotational symmetry}
If Hamilton's principle were rotational invariant under right multiplication, that is $\mathcal{L}(\chi,\dot\chi)=\mathcal{L}(\chi\chi^{-1},\dot\chi\chi^{-1})=\mathcal{L}(\Bbb{I},\dot\chi\chi^{-1})$
for any $\chi \in \mathcal{F}(\mathcal{D}, SO(3))$,
 then one could define the \emph{angular velocity matrix} 
${\color{black} \widehat{\boldsymbol{\nu}}(\mathbf{x},t)}:=
\dot\chi(\mathbf{x},t)\chi^{-1}(\mathbf{x},t) \in 
\mathfrak{so}(3)$ and the \textit{reduced Lagrangian} $\ell(\widehat{\boldsymbol{\nu}}):=
\mathcal{L}(\mathbb{I},\dot\chi\chi^{-1})$ in order to obtain the reduced \emph{Euler-Poincar\'e variational principle}
 \begin{equation}\label{fully-symmetric}
\delta\int_{t_1}^{t_2}\!\ell(\widehat{\boldsymbol{\nu}})\, 
\de^3 \mathbf{x}=0
\end{equation}
for all variations $\delta\widehat{\boldsymbol{\nu}}=
\partial_t\left((\delta\chi)\chi^{-1}\right)+
\left[(\delta\chi)\chi^{-1},\widehat{ \boldsymbol{\nu}}\right]$.

The variable $\boldsymbol{\nu} \in \mathcal{F}(\mathcal{D}, 
\mathbb{R}^3)$ is of paramount importance, first, because it is
the only variable appearing in the reduced Lagrangian $\ell$. 
Second, the physical importance of the vector $\bnu$ is that it measures the angular velocity $-(1/2)\epsilon_{ijk}\dot\chi_{ja}\chi^T_{ak}$ in the lab (\emph{spatial}) frame and thus it is a physical observable.
 
\subsection{Nematic systems and the director}\label{Sec:ELReduction}
 The hypothesis of rotational invariance leading to the Lagrangian $\ell(\widehat{ \boldsymbol{\nu}})$ is too restrictive in the context of nematic systems. This is due to the existence of a preferred  direction of each particle encoded by the director variable 
$\mathbf{n}(\mathbf{x},t)=\chi(\mathbf{x},t)\mathbf{n}_0
(\mathbf{x})$, where $\mathbf{n}_0(\mathbf{x})$ is the initial director at $t=0$.  The corresponding Lagrangian $\mathcal{L} ( \chi , \dot \chi )$ is usually written in terms of a free energy $F$, such that for nematic systems we have
 \begin{equation}\label{lagr1}
\mathcal{L}(\chi,\dot{\chi})=
\frac{J}{2}\int_\mathcal{D}\left|\dot{\chi}(\mathbf{x},t)
\mathbf{n}_0(\mathbf{x})\right|^2\,\de^3 \mathbf{x}-
\int_\mathcal{D}F\big(\chi(\mathbf{x},t)\mathbf{n}_0(\mathbf{x}),
\nabla\!\left(\chi(\mathbf{x},t)\mathbf{n}_0(\mathbf{x})\right)
\!\big)\,\de^3\mathbf{x}
\,,
\end{equation}
where $J$ is the \emph{microinertia constant}.
Notice that this $\mathcal{L}$ is \textit{not} 
$\mathcal{F}(\mathcal{D}, SO(3))$-invariant, that is, 
$\mathcal{L}(\chi,\dot\chi)\neq\mathcal{L}(\mathbb{I},
\dot\chi\chi^{-1})$. However, if we consider $\mathbf{n}_0$ as an extra variable, we obtain the following invariance property
\[
\mathcal{L}_{\bn_0}(\chi,\dot\chi)=\mathcal{L}_{\chi \bn_0}(\Bbb{I},\dot\chi\chi^{-1})\,.
\]
Thus, if we define $\bn:=\chi\bn_0$ and $\ell(\boldsymbol{\widehat\nu},\bn):=\mathcal{L}_{\chi \bn_0}(\Bbb{I},\dot\chi\chi^{-1})$, the reduced variational principle reads
\begin{equation}\label{RedHamPrinc1}
\delta\int_{t_1}^{t_2}\!\ell(\bnu,\bn)\, \de t=\int_{t_1}^{t_2}\!\left(\frac{\delta\ell}{\delta\bnu}\cdot\delta\bnu+\frac{\delta\ell}{\delta\bn}\cdot\delta\bn\right)\, \de t=0,
\end{equation}
subject to the constrained variations that are obtained by a direct
computation from the definition of the variables $\mathbf{n}$ and 
$\boldsymbol{\nu}$, namely, $\delta\mathbf{n}=
(\delta\chi)\chi^{-1}\mathbf{n}$ and $\delta\widehat{ \boldsymbol{\nu}}=\partial_t\left((\delta\chi)\chi^{-1}\right)+
\left[(\delta\chi)\chi^{-1},\widehat{\boldsymbol{\nu}}\right]$.
This yields the reduced Euler-Poincar\'e equations
\begin{equation}\label{EP-eqns1}
\left\{
\begin{array}{l}
\vspace{0.2cm}\displaystyle\frac{\partial}{\partial t}\frac{\delta \ell}{\delta \boldsymbol{\nu}}+ \frac{\delta \ell}{\delta \boldsymbol{\nu}}\times \boldsymbol{\nu}=\mathbf{n} \times \frac{\delta \ell }{\delta \mathbf{n} }\\
\displaystyle \frac{\partial\bn}{\partial t} + \mathbf{n} \times \boldsymbol{\nu }=0.
\end{array}\right.
\end{equation}
Notice that the system is fully $SO(3)$-invariant if and only if $\bn=0$; otherwise one says the $SO(3)$ symmetry is \emph{broken} by the presence of the special direction $\mathbf{n}_0$. 

The above equations provide the convenient setting for the derivation of the Ericksen-Leslie equation by taking the Lagrangian
\begin{equation}\label{EL-Lagr}
\ell( \boldsymbol{\nu},  \mathbf{n} )= 
\frac{J}{2} \int_{\mathcal{D}}\left|\boldsymbol\nu\times\mathbf{n} \right|^2 \,\de^3 \mathbf{x} -
\int_{\mathcal{D}}F(\mathbf{n},\nabla\mathbf{n}) \,\de^3 \mathbf{x}\,,
\end{equation}
computing its variational derivatives
\begin{align*}
\frac{\delta \ell}{\delta\bnu}=-J\,\mathbf{n}\times(\mathbf{n}\times \boldsymbol{\nu})=J\,\mathbf{n}\times\partial_t\bn
\,,\qquad
\frac{\delta \ell}{\delta\bn}=-J\,\boldsymbol{\nu}\times(\boldsymbol{\nu}\times\mathbf{n})-\mathbf{h}=-J\,\boldsymbol{\nu}\times\partial_t\bn-\mathbf{h},
\end{align*}
and applying $\mathbf{n}\times$ to the $\boldsymbol{\nu}$-equation 
after using the Jacobi identity.

\subsection{Eringen's wryness tensor}
As we have seen, the free energy contains the derivatives of 
$\mathbf{n}$. This is not a problem because $\nabla\mathbf{n}$  
depends  on the director $\mathbf{n}$, so that the explicit dependence of $F$ on $\nabla\bn$ is irrelevant. For example, the molecular field is expressed as
\[
\mathbf{h}=\frac{\delta F}{\delta \bn}-\frac{\partial}{\partial x^i}\frac{\delta F}{\delta (\partial_{x^i}\bn)}
\,.
\]

However, we want to emphasize that the spatial derivatives of 
$\mathbf{n}$ can be used to introduce an extra variable, as follows. Since
\[
\nabla\mathbf{n}=\nabla(\chi\mathbf{n}_0)=
(\nabla\chi)\chi^{-1}\mathbf{n}+\chi\nabla\mathbf{n}_0\,,
\]
it follows that if $\mathbf{n}_0$ is spatially constant, so 
$\nabla\mathbf{n}_0=0$, one can use the $\mathfrak{so}(3)$-valued
one form $\widehat{\boldsymbol{\gamma}}
:=-(\nabla\chi)\chi^{-1}$ on $\mathcal{D} \subset \mathbb{R}^3$  to write $\partial_i\mathbf{n}=\mathbf{n}\times\boldsymbol{\gamma}_i$, where the components of $\boldsymbol{\gamma}$ are 
$\gamma_i^a=-(1/2)\epsilon^{abc\,}\widehat\gamma_{i}^{\,bc}$. If $\nabla\bn_0\neq0$, then there exist three non vanishing vectors $\bgamma_{0\,i}$ such that $\partial_i\bn_0=\bn_0\times\bgamma_{0\,i}$. This can be seen by recalling that the condition $| \mathbf{n} _0 (x)|^2=1$ yields $\partial_i \mathbf{n}_0 \cdot \mathbf{n}_0  =0$, so that the three vectors 
\[
\bgamma_{0\,i}=\partial_i\bn_0\times\bn_0+\lambda_i\bn_0
\]
are defined up to an additive vector parallel to $\bn_0$. Then, from the previous equation, we obtain
\[
\nabla\mathbf{n}=\nabla(\chi\mathbf{n}_0)=
\left((\nabla\chi)\chi^{-1}-
\chi\widehat{\boldsymbol{\gamma}}_0\chi^{-1}\right)\mathbf{n}=:
\mathbf{n}\times\boldsymbol{\gamma}
\]
where we have now defined
\begin{equation}
\label{gamma_def}
\widehat{\boldsymbol{\gamma}}:=-(\nabla\chi)\chi^{-1}+
\chi\widehat{ \boldsymbol{\gamma}}_0\chi^{-1}=
\chi(\nabla\chi^{-1})+\chi\widehat{\boldsymbol{\gamma}}_0\chi^{-1}.
\end{equation}
Following Eringen's work, we call the tensor $\boldsymbol{\gamma}$ 
with components $\gamma_i^a$ the \emph{wryness tensor}. It is 
important to notice that we sharply diverge here with Eringen's 
considerations, who defined $\boldsymbol{\gamma}= (\nabla\mathbf{n})\times\mathbf{n}$, a quantity that is \textit{not} gauge invariant
and would require $\boldsymbol{\gamma}_i \cdot \mathbf{n} = 0$ at
all times. As we shall see below, it is precisely our definition 
of $\boldsymbol{\gamma}$ that will lead to the main result:  
Eringen's micropolar theory contains EL nematodynamics as a special 
case. 

Returning to our definition \eqref{gamma_def} of 
$\boldsymbol{\gamma}$ in terms of $\chi$, we notice that the initial condition $\widehat{\boldsymbol{\gamma}}_0$ appears precisely as a \emph{gauge transformation} by $\chi$ applied to the (Yang-Mills) magnetic potential $\widehat{ \boldsymbol{\gamma}}_0$. This fact leads us to consider a nematic Lagrangian 
$\mathcal{L}(\chi,\dot\chi)$ of the type
\begin{align}\label{primitive_Lagr} 
\mathcal{L}(\chi,\dot{\chi})&=
\frac{J}{2}\int_\mathcal{D}\left|\dot\chi(\mathbf{x},t)
\mathbf{n}_0(\mathbf{x})\right|^2\,\de^3 \mathbf{x}\nonumber\\
& \qquad -
\int_\mathcal{D}F\big(\chi(\mathbf{x},t)\mathbf{n}_0(\mathbf{x}),
(\nabla\chi(\mathbf{x},t)- \chi(\mathbf{x},t)
\widehat{\boldsymbol{\gamma}}_0(\mathbf{x}))
\mathbf{n}_0(\mathbf{x})\big)\,\de^3\mathbf{x}
\,.
\end{align}
Again, the $SO(3)$-invariance is lost by the above total Lagrangian, that is $\mathcal{L}(\chi,\dot\chi)\neq\mathcal{L}(\Bbb{I},\dot\chi\chi^{-1})$. However, if we consider $\mathbf{n}_0$ and 
$\widehat{\boldsymbol{\gamma}}_0$ as  extra variables, we  obtain the following invariance property
\[
\mathcal{L}_{\mathbf{n}_0,\widehat{\boldsymbol{\gamma}}_0}
(\chi,\dot{\chi})=
\mathcal{L}_{\mathbf{n},\widehat{\boldsymbol{\gamma}}}(\mathbb{I},\dot\chi\chi^{-1})
\,,
\]
where 
\begin{equation}\label{def_n_gamma} 
(\bn,\widehat\gamma)=
\left(\chi\mathbf{n}_0,\chi(\nabla\chi^{-1})+
\chi\widehat{\boldsymbol{\gamma}}_0\chi^{-1}\right)
\,.
\end{equation} 
 These definitions define an \emph{action} of the gauge group $\mathcal{F(D},SO(3))$ on the Cartesian product ${\mathcal{F(D},S^2)\times\Omega^1(\mathcal{D},\mathfrak{so}(3))}$, where $\mathcal{F(D},S^2)$ is the space of director fields while $\Omega^1(\mathcal{D},\mathfrak{so}(3))$ denotes the space $\Omega^1(\mathcal{D})\otimes\mathfrak{so}(3)$ (containing $\boldsymbol{\widehat\gamma}$) of $\mathfrak{so}(3)$-valued one forms on $\mathcal{D}$. See e.g., \cite{MaRa2004} for
basic information on group actions.  In this particular case, the 
$\chi$-action $\chi\mathbf{n}_0$ on the director $\mathbf{n}_0$ is linear, i.e., it is a representation. On the other hand, the 
$\chi$-action $\chi(\nabla\chi^{-1})+
\chi\widehat{\boldsymbol{\gamma}}_0\chi^{-1}$ on the magnetic potential $\boldsymbol{\widehat\gamma}_0$ is an \emph{affine action}; see \cite{GBRa2009}. Notice that these group actions emerged \emph{naturally} from the problem and the concept of a group action was not required at the beginning of our discussion, which started with a Lagrangian depending on the parameters $\bn_0$ and $\bgamma_0$. 

{\color{black} Proceeding as in the previous case, we introduce the reduced Lagrangian $\ell(\widehat{\boldsymbol{\nu}},\mathbf{n},
\widehat{\boldsymbol{\gamma}}):= 
\mathcal{L}_{\mathbf{n},\widehat{\boldsymbol{\gamma}}}(\mathbb{I},\dot\chi\chi^{-1})$ arising from the primitive Lagrangian $\mathcal{L}_{\mathbf{n}_0,\widehat{\boldsymbol{\gamma}}_0}
(\chi,\dot{\chi})$ in \eqref{primitive_Lagr}, and get the reduced variational principle 
\begin{equation}\label{VarPrinc-gamma}
\delta\int_{t_1}^{t_2}\!\ell(\bnu,\bn,\bgamma)\, \de^3 x=\int_{t_1}^{t_2}\!\left(\frac{\delta\ell}{\delta\bnu}\cdot\delta\bnu+\frac{\delta\ell}{\delta\bn}\cdot\delta\bn+\frac{\delta\ell}{\delta\bgamma_i}\cdot\delta\bgamma_i\right)=0
\end{equation}
subject to the constrained variations 
\begin{align*}
\left(\delta\mathbf{n},\delta\widehat{\boldsymbol{\gamma}}\right)
&=\left((\delta\chi)\chi^{-1}\mathbf{n},
-\nabla\left((\delta\chi)\chi^{-1}\right)-
\left[\widehat{\boldsymbol{\gamma}},(\delta\chi)\chi^{-1}\right]\right), \\
\delta\widehat{\boldsymbol{\nu}}&=
\partial_t\left((\delta\chi)\chi^{-1}\right)+
\left[(\delta\chi)\chi^{-1},\widehat{\boldsymbol{\nu}}\right],
\end{align*}
obtained  by a direct computation from the definition of the variables 
$\widehat{\boldsymbol{\nu}} = \dot \chi \chi ^{-1} $, $ \mathbf{n} $, and $ \widehat{\boldsymbol{\gamma}}$ given in \eqref{def_n_gamma}.
The variational principle \eqref{VarPrinc-gamma} yields the reduced Euler-Poincar\'e equations
\begin{equation*}
\left\{
\begin{array}{l}
\vspace{0.2cm}\displaystyle\frac{d}{dt}\frac{\delta \ell }{\delta \boldsymbol{\nu}}=\boldsymbol{\nu}\times \frac{\delta \ell }{\delta \boldsymbol{\nu}}+\frac{\partial}{\partial x^i}\frac{\delta\ell_2}{\delta\bgamma_i}+ \boldsymbol{\gamma }_i \times \frac{\delta \ell }{\delta \boldsymbol{\gamma}_i }   +\mathbf{n} \times \frac{\delta \ell  }{\delta \mathbf{n} }\\
\vspace{0.2cm}\displaystyle \partial_ t \mathbf{n} + \mathbf{n} \times \boldsymbol{\nu }=0\\
\displaystyle \partial _t \bgamma_i +  \bgamma_i \times \boldsymbol{\nu}+ \partial_i \boldsymbol{\nu}=0,
\end{array}\right.
\end{equation*}
which are equivalent to the Euler-Lagrange equations for the primitive Lagrangian $\mathcal{L}_{\mathbf{n}_0,\widehat{\boldsymbol{\gamma}}_0}
(\chi,\dot{\chi})$.} 
These equations provide an alternative description of nematic systems. When $\boldsymbol{\gamma}_0=0$ and  
$\nabla\mathbf{n}_0=0$, the above system is fully equivalent to 
that obtained in the previous section. This follows from the fact 
that the  Lagrangian $\mathcal{L}(\chi,\dot\chi)$ in material 
representation has never been modified and the only change was 
introduced through the initial condition $\nabla\mathbf{n}_0=
\mathbf{n}_0\times\boldsymbol{\gamma}_0$, which led to the 
introduction of $\boldsymbol{\gamma}$. For example, upon replacing 
$\nabla\mathbf{n}=\mathbf{n}\times\boldsymbol{\gamma}$ (with 
$\boldsymbol{\gamma}_0=0$) in the Frank energy we obtain a free 
energy $\Phi(\mathbf{n},\boldsymbol{\gamma})$ and the variational 
derivatives
\[
\frac{\delta \Phi}{\delta\boldsymbol{\gamma}_i}
=-\mathbf{n}\times\frac{\delta F}{\delta(\partial_i\mathbf{n})}
\,,\qquad
\frac{\delta \Phi}{\delta\mathbf{n}}
=\frac{\delta F}{\delta\mathbf{n}}+
\boldsymbol{\gamma}_i\times
\frac{\delta F}{\delta(\partial_i\mathbf{n})}
\]
to produce
\[
\mathbf{n}\times\mathbf{h}=
\partial_i\frac{\delta \Phi}{\delta\boldsymbol{\gamma}_i}+
\boldsymbol{\gamma}_i\times
\frac{\delta \Phi}{\delta\boldsymbol{\gamma}_i}+
\mathbf{n}\times\frac{\delta \Phi}{\delta\mathbf{n}}\,,
\]
which leads to the compatibility between the two approaches.

\begin{remark}[Eringen's formulation of the wryness tensor]{\rm 
The explicit form of the wryness tensor lies at the origin of many problems that were encountered to show compatibility of Ericksen-Leslie theory with Eringen's micropolar theory. In particular, equation \eqref{gamma_def} was well known to Eringen (at least in the case when $\widehat{\bgamma}_0=0$), who first defined the wryness tensor as $\widehat{\bgamma}=-(\nabla\chi)\chi^{-1}$. In \cite{Eringen1993}, he noticed that  $\bgamma$ cannot be solved uniquely from the equation $\nabla\bn=\bn\times\bgamma$ and the identification $\bgamma=\nabla\bn\times\bn$ requires $\bgamma\cdot\bn=0$. However, in the same paper, it is claimed that the non-uniqueness of ${\bgamma}$ does not affect the expression of the free energy $\Phi$, so that one can still use the relation $\bgamma=\nabla\bn\times\bn$. This statement is contradicted by the fact that the condition $\bgamma\cdot\bn=0$ is \emph{not} preserved in time and thus the quantity $\nabla\bn\times\bn$ cannot be identified with the dynamical variable $\bgamma$ at all times. This point has been at the origin of all problems concerning Eringen's micropolar theory; problems that are now solved in the present paper, which shows how the micropolar theory extends Ericksen-Leslie nematodynamics by allowing for an arbitrary initial condition 
$\bgamma_0$.}
\end{remark}
 However, when $\bgamma_0\neq0$, the relation between Ericksen-Leslie dynamics and Eringen's micropolar model is more subtle. Notice that the micropolar dynamics yields
\[
\left(\frac{\partial}{\partial t}-\bnu\times\right)\left(\nabla\bn-\bn\times\bgamma\right)=0
\]
where $\bn=\chi\bn_0$ and $\bgamma=-(\nabla\chi)\chi^{-1}+
\chi\widehat{ \boldsymbol{\gamma}}_0\chi^{-1} $ are independent dynamical variables. In turn, this implies
\[
\nabla\bn-\bn\times\bgamma=\chi\left(\nabla\bn_0-\bn_0\times\bgamma_0\right)
.
\]
Then, if the initial conditions satisfy $\nabla\bn_0=\bn_0\times\bgamma_0$, this relation is preserved by the dynamics and Eringen's micropolar theory is still equivalent to Ericksen-Leslie nematodynamics. On the other hand, when the initial conditions of micropolar dynamics are such that $\nabla\bn_0\neq\bn_0\times\bgamma_0$, then the two theories are not compatible. Notice that, when $\bgamma_0\neq0$, a non vanishing magnetic Yang-Mills field $\boldsymbol{B}_0^i=\epsilon^{ijk}\!\left(\partial_j\bgamma_{0\,k}+\bgamma_{0\,j}\times\bgamma_{0\,k}\right)$ defines the (initial) \emph{disclination density} in the gauge theory of disclinations  \cite{Volovick1980}. If the initial conditions are such that $\boldsymbol{B}_0^i=0$ (e.g., $\bgamma_{0\,i}=0$), then the disclination density vanishes at all times \cite{GBRa2009}.

%%%%%%%%%%%%%%%%%%%%%%%%%%%%%%%%%%%%%%%%%%%%%%%%%%%%%%%%%%%%%%%%%%%%%%%%%%%%%

\section{Broken symmetry and Noether's theorem}

\subsection{Residual symmetries and isotropy subgroups}
In this section, we explain how the various ways in which symmetry 
can be used lead to different conserved quantities arising from 
Noether's theorem. These conserved quantities arise from the fact 
that, although the full $SO(3)$ symmetry is broken in nematic systems 
(i.e., $\mathcal{L}(\chi,\dot\chi)\neq\mathcal{L}(\mathbb{I},
\dot{\chi}\chi^{-1})$), \emph{residual symmetries} are still 
possible. This means that there exists a Lie subgroup $G_0$ of 
$\mathcal{F}(\mathcal{D},SO(3)$ such that $\mathcal{L}(\chi,\dot\chi)
=\mathcal{L}(\chi\chi_0^{-1},\dot\chi\chi_0^{-1})$ for any 
$\chi_0\in G_0$. This Lie subgroup leaving $\mathcal{L}$ invariant 
is known as \emph{isotropy subgroup} and it is associated to the 
variable breaking the symmetry, e.g., the director $\mathbf{n}_0$. 
In particular, the isotropy subgroup of $\mathbf{n}_0$ is defined as
\[
G_{\bn_0}=\left\{\chi_0\in \mathcal{F(D},SO(3))\mid
\chi_0\mathbf{n}_0=\mathbf{n}_0\right\}.
\]
One verifies immediately that the Lagrangian \eqref{lagr1}  is invariant under $G_{\bn_0}$. For simplicity, here we consider the case in which $\bn_0=\mathbf{e}_3$ so that $\nabla\bn_0=0$. Then, 
\[
G_{\mathbf{e}_3}=\mathcal{F(D},SO(2))
\,,
\]
the gauge group of planar rotations,
Notice that, strictly speaking, one should make use of the 
(\emph{dihedral}) group ${\mathcal{F(D},SO(2))\times\Bbb{Z}_2}$ to 
account for the equivalence $\bn_0\sim-\bn_0$; however, this reflection 
symmetry is left as an \emph{a posteriori} verification in this 
treatment. {\color{black} This is consistent with most of liquid crystal theories. In 
principle, the discrete $\mathbb{Z}_2$-symmetry could also be taken into 
account by using sophisticated methods in reduction theory 
\cite{MaMiOrPeRa2007}. However, the use of these methods would require 
a much more involved treatment, which is left for future work.}

By proceeding analogously, one can also define the isotropy subgroup of $\bgamma_0$, upon using the affine $\chi$-action introduced previously, to get
\[
G_{\widehat{\boldsymbol{\gamma}}_0}=
\left\{\chi_0\in \mathcal{F(D},SO(3))\mid 
\chi_0\widehat{\boldsymbol{\gamma}}_0\chi_0^{-1}+
\chi_0(\nabla\chi_0^{-1})=\widehat{\boldsymbol{\gamma}}_0\right\}\,.
\]
In the spatial case when $\widehat{\boldsymbol{\gamma}}_0=0$ 
(i.e., $\nabla\mathbf{n}_0=0$), its isotropy subgroup is
\[
G_{\widehat{\boldsymbol{\gamma}}_0=0}=SO(3)
\,,
\]
that is, the Lie group of spatially constant rotation matrices 
$\chi_0\in SO(3)$. In more generality, the isotropy subgroup of the couple $(\mathbf{e}_3,\widehat{\boldsymbol{\gamma}}_0=0)$ is given 
by $G_{\mathbf{e}_3}\cap G_{\widehat{\boldsymbol{\gamma}}_0=0}=
\mathcal{F(D},SO(2))\cap SO(3)=SO(2)$, i.e., the Lie group of spatially constant planar rotations.

\subsection{Geometric nature of conserved quantities}

{\color{black} Noether's theorem ensures} that there are conserved 
quantities corresponding to each of the previous symmetry subgroups. 
The reduced equations provide a convenient setting for expressing 
these conserved quantities. In the simplest case of a fully 
$\mathcal{F} (\mathcal{D} ,SO(3))$-invariant system, one has $\mathbf{n}_0=0$ (i.e.,
$\mathbf{n}=\chi\mathbf{n}_0=0$) and the associated conserved quantity is {\color{black} $\chi^{-1}(\delta\ell/\delta\boldsymbol{\nu})$, that is,}
\[
\frac{\de}{\de t}\left(\chi^{-1}
\frac{\delta \ell}{\delta \boldsymbol{\nu}}\right)
=\chi^{-1}\!\left(\frac{\de}{\de t}
\frac{\delta \ell}{\delta \boldsymbol{\nu}}+
\frac{\delta \ell}{\delta \boldsymbol{\nu}}\times\boldsymbol{\nu}
\right)=0,
\]
where $\ell(\boldsymbol{\nu})$ is obtained from the Lagrangian 
$\ell(\widehat{\boldsymbol{\nu}})$ appearing in 
\eqref{fully-symmetric} by using the hat map.
The above relation is the specialization of a well known result in 
the geometric mechanics of invariant systems \cite{MaRa2004}. Notice 
that {\color{black} the momentum variable $\delta\ell/\delta\boldsymbol{\nu}$ belongs to
$\mathfrak{so}(3)^*$, the dual space of the Lie algebra 
$\mathfrak{so}(3)\simeq\mathbb{R}^3$, and so does $\chi^{-1}\delta\ell/\delta\boldsymbol{\nu}$ (since the latter identifies  the group coadjoint action $\operatorname{Ad}_{\chi^{-1}}\delta\ell/\delta\boldsymbol{\nu}$). Thus one again identifies 
$\mathfrak{so}(3)^*\simeq\mathbb{R}^3$. 
\begin{framed}\it\noindent
As a general principle in 
geometric mechanics, conserved quantities belong to the dual space 
of the Lie algebra associated to the symmetry group leaving the 
system invariant.
\end{framed}}
\noindent
This means that if Hamilton's principle is 
invariant under the action of a Lie group $G$, then Noether's 
conserved quantity belongs to the dual space $\mathfrak{g}^*$ of 
the Lie algebra $\mathfrak{g}$ of $G$.

At this point we wonder what happens when the rotational symmetry 
is broken (i.e., $\mathbf{n}=\chi\mathbf{n}_0\neq0$) and the above conservation transforms into the relation
\[
\frac{\de}{\de t}\left(\chi^{-1}\frac{\delta \ell}{\delta \bnu}\right)=\chi^{-1}\!\left(\frac{\de}{\de t}
\frac{\delta \ell}{\delta \boldsymbol{\nu}}+
\frac{\delta \ell}{\delta \boldsymbol{\nu}}\times\boldsymbol{\nu}
\right)=\chi^{-1}\!\left(\bn\times\frac{\delta \ell}{\delta \bn}\right)
\,,
\]
which arises from the first equation in \eqref{EP-eqns1}.

Following the general principle just stated, we seek a conserved 
quantity belonging to the dual  Lie algebra of the isotropy group 
$G_{\mathbf{n}_0}$, which leaves \eqref{lagr1} invariant. Upon 
specializing to the case $\mathbf{n}_0=\mathbf{e}_3$, we seek 
a quantity in the dual space of $\mathfrak{g}_{\mathbf{e}_3}=
\mathcal{F(D},\mathfrak{so}(2))=C^\infty(\mathcal{D})$. A  
geometrically natural method to construct such quantity is to use 
the dual of the \emph{Lie algebra inclusion} $\boldsymbol{i}(r)=
(0,0,r)\in\mathcal{F(D},\mathbb{R}^3)\simeq\mathcal{F(D},
\mathfrak{so}(3))$ for an arbitrary scalar function 
$r\in C^\infty(\mathcal{D})\simeq\mathcal{F(D},\mathfrak{so}(2))$. 
The dual operator $\boldsymbol{i}^*$ is defined as 
${\int_\mathcal{D}\boldsymbol\mu\cdot\boldsymbol{i}(r)\,\de^3 x}
=:{\int_\mathcal{D}r\boldsymbol{i}^*(\boldsymbol\mu)\,\de^3x}$ for 
an arbitrary vector function $\boldsymbol\mu\in
\mathcal{F(D},\mathfrak{so}(3))^*\simeq\mathcal{F(D},\mathbb{R}^3)$. 
Thus, $\boldsymbol{i}^*(\boldsymbol\mu)=\mu_3$, the third component 
of $\boldsymbol\mu$, and we obtain
\[
\boldsymbol{i}^*\!\left(\chi^{-1}
\frac{\delta \ell}{\delta\boldsymbol{\nu}}\right)=
\left(\chi^{-1}\frac{\delta \ell}{\delta\boldsymbol{\nu}}
\right)_{\!3}=\chi_{3i}^T\frac{\delta \ell}{\delta\boldsymbol{\nu}_i}
= \mathbf{n}\cdot\frac{\delta \ell}{\delta\boldsymbol{\nu}}\,,
\]
where we have used Einstein's summation convention and the fact 
that $\mathbf{n}_l=\chi_{lk}\mathbf{n}_{0k}=\chi_{lk}\delta_{k3}=
\chi_{l3}$. In conclusion, the reduced Hamilton's principle 
\eqref{RedHamPrinc1} is accompanied by the conservation law
\begin{equation}\label{cons_law_1} 
\frac{\de}{\de t}\!\left(\mathbf{n}\cdot
\frac{\delta \ell}{\delta \boldsymbol{\nu}}\right)=0
\,,
\end{equation} 
as can be easily verified from the Euler-Poincar\'e equations 
\eqref{EP-eqns1}.

When the wryness tensor is also considered as a dynamical variable, 
then one seeks a conserved quantity belonging to the dual Lie algebra 
of the isotropy Lie subgroup $G_{\mathbf{e}_3,\bgamma_0=0}=SO(2)$.  
Using the same method as above, one considers the Lie algebra 
inclusion $\boldsymbol{i}(r)=(0,0,r)\in\mathcal{F(D},
\mathbb{R}^3)\simeq\mathcal{F(D},\mathfrak{so}(3))$ for any real 
number $r\in\mathfrak{so}(2)$. Then, the corresponding dual operator is 
$\boldsymbol{i}^*(\boldsymbol\mu)=\int_\mathcal{D}\mu_3(\bx)\,\de^3 x$ 
and the conserved quantity associated to the reduced Hamilton's 
principle \eqref{VarPrinc-gamma} is {\color{black} $\int_\mathcal{D\,}\bn\cdot
{\delta \ell}/{\delta \bnu}\,\de^3 \mathbf{x}$ (see 
\cite{GBRaTr2011}), i.e.,}
\begin{equation}\label{cons_law_2} 
\frac{\de}{\de t}\int_\mathcal{D}\bn\cdot\frac{\delta \ell}{\delta \bnu}
\,\de^3 \mathbf{x}=0
\,,
\end{equation} 
as long as $\bn_0=\mathbf{e}_3$ and $\bgamma_0=0$.

\subsection{Conserved quantities are momentum maps}\label{cons_quant}
In geometric mechanics, Noether's conservation laws are reformulated 
in terms of \emph{momentum maps}. These maps take values in the dual 
of a Lie algebra and always identify the conserved quantities of 
symmetric systems, {\color{black}i.e., systems whose Hamilton's 
principle is invariant under the action of a Lie group $G_0$. The 
various (and important) properties of momentum maps assume some 
basic knowledge of Lie group theory, which can be found in geometric 
mechanics books (e.g., \cite{MaRa2004}). 
One of the most important results in mechanics linking 
symmetry to conservation laws states that 

\medskip
{\it\noindent
momentum maps always identify Noether's conserved quantities when they 
arise from the same symmetry group that leaves Hamilton's principle 
invariant.}

\medskip
\noindent
The mechanical system under consideration is \textit{symmetric} if 
its Lagrangian $\mathcal{L}(\chi,\dot\chi)$ is invariant with respect 
to the action of a certain Lie group $G_0$ on $TQ$, i.e., 
$\mathcal{L}(\chi,\dot\chi)=\mathcal{L}(g_0\chi,g_0\dot\chi)$ for any 
$g_0\in G_0$. In the special case when the $G_0$-action on the tangent 
bundle $TQ$ is the  lift of a $G_0$-action on  $Q$ (see \cite{MaRa2004}),
the momentum map $J:T^*Q\to\mathfrak{g}_0^*$, where  $T^*Q$ 
is  the cotangent bundle of $Q$ and $\mathfrak{g}_0$ is the Lie 
algebra of $G_0$, is given by
\begin{equation}\label{momap-formula}
\left\langle\!\left\langle J\!\left( \chi , \mathcal{P} \right)\!, 
\xi \right\rangle\!\right\rangle = \left\langle \mathcal{P} , 
\xi _Q( \chi ) \right\rangle,
\end{equation}
for $\mathcal{P} := \frac{\delta\mathcal{L}}{\delta\dot\chi}$ the 
momentum conjugate to the microvelocity $\dot\chi$. In formula 
\eqref{momap-formula}, $\left\langle\!\left\langle \cdot,\cdot 
\right\rangle\!\right\rangle$ denotes the pairing between vectors in 
$\mathfrak{g}_0$ and their duals in $\mathfrak{g}_0^\ast $, 
$\left\langle \cdot,\cdot\right\rangle$ 
is the  vector-covector pairing on $Q$, and $\xi_Q(\chi)$ denotes the 
action of the \emph{infinitesimal generator} vector field $\xi_Q \in 
\mathfrak{X}(Q)$ defined by $\xi\in \mathfrak{g}_0$ (corresponding to 
the $G_0$-action on $Q$) on $\chi\in Q$. The concept of infinitesimal 
generator is of paramount importance in geometric mechanics and it 
is defined as the derivative of the group action. For example, the 
infinitesimal generator of the $SO(3)$-action on the initial 
director field $\mathbf{n}_0\in S^2$ is
\[
\left.\frac{\de}{\de t}\right|_{t=0}
\!\left(\chi(t)\mathbf{n}_0\right)
=\dot\chi(0)\mathbf{n}_0
=\boldsymbol{\nu}_0\times\mathbf{n}_0
=\nu_{S^2}(\bn_0),
%=\dot{\mathbf{n}}_0(\mathbf{x})
\]
where $ \widehat{\boldsymbol{\nu}_0}=\dot\chi(0)$, $\chi(0)=\Bbb{I}$, 
and we  omitted spatial dependence.

In the context of Ericksen-Leslie nematodynamics, the configuration manifold is $\mathcal{F(D},SO(3))$ and the symmetry group is the subgroup $G_0\subset\mathcal{F(D},SO(3))$ that leaves $\bn_0$ invariant, that is, the group of planar rotations on the plane perpendicular to $\bn_0$. Upon recalling the spatial dependence  $\bn_0= \bn_0(\mathbf{x})$, we identify $G_0=\mathcal{F(D},SO(2))$. This is a typical situation in which the (gauge) $SO(3)$ symmetry is broken by the $SO(2)$ (gauge) group. In more generality, systems with broken symmetries involve a Lie group configuration manifold $Q=G$ and a symmetry subgroup $G_0\subset G$. Then, at the Lie algebra level, there is a subgroup inclusion $i:\mathfrak{g}_0\hookrightarrow\mathfrak{g}$ so that the infinitesimal generator of the $G_0$-action on $G$ is induced by the infinitesimal generator of the action of $G$ on itself. Upon fixing $\xi\in\mathfrak{g}_0$, we have
\[
\xi_G(\chi)=\left(i(\xi)\right)_G\chi
\]
and on the right hand side we consider the infinitesimal generator of the right multiplication in $G$. Then, $\left(i(\xi)\right)_G( \chi )=\chi\,i(\xi)$ and, by denoting $\mathcal{P}=\delta\mathcal{L}/\delta\dot\chi$, the momentum map formula \eqref{momap-formula} gives 
\[
\left\langle\!\left\langle J( \chi , \mathcal{P} ), \xi \right\rangle\!\right\rangle = \left\langle \mathcal{P} , \left(i(\xi)\right)_G( \chi ) \right\rangle
=
\left\langle \mathcal{P} ,\chi\,i(\xi) \right\rangle
=
\left\langle\!\left\langle \chi^{-1}\mathcal{P} ,i(\xi) \right\rangle\!\right\rangle
=
\left\langle\!\left\langle i^*\!\!\left(\chi^{-1}\mathcal{P}\right) ,\xi \right\rangle\!\right\rangle
\]
where the third step is due to the invariance of the pairing and the notation $\left\langle\!\left\langle \cdot,\cdot \right\rangle\!\right\rangle$ stands for the pairing on $\mathfrak{g}^*\times\mathfrak{g}$ or $\mathfrak{g}_0^*\times\mathfrak{g}_0$, depending on the context. The momentum map is therefore
\[
J\!\left(\chi,\frac{\delta\mathcal{L}}{\delta\dot\chi}\right)=i^*\!\left(\chi^{-1}\frac{\delta\mathcal{L}}{\delta\dot\chi}\right)
\]
and upon using the right trivialization $(\chi,\dot\chi)\mapsto(\chi,\nu)$ (with $\nu=\dot\chi\chi^{-1}$), the above formula is rewritten in terms of the reduced Lagrangian as
\begin{equation}\label{Noether-quantity}
J\!\left(\chi,\frac{\delta\ell}{\delta\nu}\right)=i^*\!\left(\operatorname{Ad}^*_\chi\frac{\delta\ell}{\delta\nu}\right)
= i^*\!\left(\chi^{-1}\frac{\delta\ell}{\delta\nu}\chi\right), 
\end{equation}
where $\operatorname{Ad}^*$ stands for the group coadjoint operation on $G$; see e.g., \cite{MaRa2004}. The latter formula  provides Noether's conserved quantity for an arbitrary system with broken symmetry.

When the above momentum map is specialized to the case of Ericksen-Leslie nematodynamics, it is convenient to assume $\bn_0=\mathbf{e}_3$ so that the Lie algebra inclusion $i:C^\infty(\mathcal{D})\hookrightarrow\mathcal{F}(\mathcal{D},\Bbb{R}^3)$ is $i(r(\mathbf{x}))=(0,0,r(\mathbf{x}))^T$ and the momentum map determines the conservation law \eqref{cons_law_1}. On the other hand, when the wryness tensor $\bgamma$ is introduced, the choice $\bgamma_0=0$ leads to the subgroup $G_0=SO(2)\subset\mathcal{F(D},SO(3))$, thereby leading to the conserved integral in \eqref{cons_law_2} .
}

\section{Euler-Poincar\'e reduction for micropolar theory}

Nematic systems are usually described by the Ericksen-Leslie theory. 
However, this theory limits itself to describe systems of uniaxial 
molecules in the absence of orientational defects (disclinations). As 
we saw, disclinations can be incorporated into the Ericksen-Leslie 
description by assuming an initial configuration 
$\boldsymbol{\gamma}_0\neq0$. However, the resulting description in 
terms of the director $\mathbf{n}$ still provides no information 
about other possible molecule conformations, which can be for example 
biaxial. Eringen's micropolar theory \cite{Eringen1997} accounts for 
more general molecule shapes by attaching to each particle a 
\emph{microinertia tensor} $j$. The resulting description has to be 
consistent with the Ericksen-Leslie theory for uniaxial molecules and 
this has always been an issue in liquid crystal modeling. As we shall 
see, the two theories are indeed consistent and, more particularly, 
Eringen's theory reduces to Ericksen-Leslie for the case of uniaxial 
nematic molecules in the absence of disclinations.

\subsection{The Lhuillier-Rey Lagrangian}

In order to insert a microinertia tensor in the dynamics, let us 
rewrite the primitive Ericksen-Leslie Lagrangian \eqref{EL-Lagr} as
\begin{align}\label{lagr-Eringen}
\ell(\boldsymbol\nu,\bn)= 
\frac{J}{2} \int_{\mathcal{D}}\big(\left|\boldsymbol\nu\right|^2-(\boldsymbol\nu\cdot\bn)^2\big)\, \de^3 \mathbf{x} -
\int_{\mathcal{D}}F(\mathbf{n},\nabla\mathbf{n}) \,\de^3 \mathbf{x}\,,
\end{align}
and notice that, if we introduce the tensor parameter  
$j(\mathbf{x},t):=J(\mathbb{I}-\mathbf{n}(\mathbf{x},t)
\mathbf{n}^T(\mathbf{x},t))$, we obtain a new Lagrangian
\begin{equation}\label{lagr-LR}
\ell(\boldsymbol\nu,j,\bn)=
\frac{J}{2} \int_{\mathcal{D}}\boldsymbol\nu\cdot j\boldsymbol\nu \,\de^3 \mathbf{x} -
\int_{\mathcal{D}}F(\mathbf{n},\nabla\mathbf{n}) \,\de^3 \mathbf{x}\,,
\end{equation}
where the microinertia tensor $j(\bx,t)$ evolves according to $j(\bx,t)=\chi(\bx,t) j_0(\bx)\chi^{-1}(\bx,t)$, with $j_0(\mathbf{x}):=J(\mathbb{I}-\mathbf{n}_0(\mathbf{x})\mathbf{n}_0^T(\mathbf{x}))$. Then, upon making use of the variation $\delta j=\left[\delta\chi\chi^{-1},j\right]$, the Euler-Poincar\'e variational principle
\[
\delta\int^{t_2}_{t_1}\ell(\boldsymbol\nu,j,\bn)\,\de t=0
\]
yields the 
equations of motion
\begin{equation}\label{EP-LR1}
\left\{
\begin{array}{l}
\vspace{0.2cm}
\displaystyle\frac{\partial}{\partial t}
\frac{\delta \ell}{\delta \boldsymbol{\nu}}=
\boldsymbol{\nu}\times \frac{\delta \ell}{\delta \boldsymbol{\nu}}
+\!\overrightarrow{\,\left[j,\frac{\delta \ell}{\delta j}\right]\,}
+\mathbf{n} \times \frac{\delta \ell}{\delta \mathbf{n} }\\
\vspace{0.2cm}\displaystyle\partial_t j+[j,\widehat{{\nu}}]=0\\
\displaystyle \partial_ t \mathbf{n} + \mathbf{n} \times 
\boldsymbol{\nu }=0,
\end{array}\right.
\end{equation}
where we have introduced the notation $\overrightarrow{A}_l=
\epsilon_{lhk}A_{hk}$. After some simplifications, using the 
expression of $\ell$, the first equation becomes
\[
\frac{\partial}{\partial t}(j\bnu)=\mathbf{h}\times\bn
\,.
\]
Then, upon allowing for more general forms of the microinertia $j$ (which then becomes an independent variable), the above equations form the Lhuiller-Rey model \cite{LR} for continuum media. This model evidently reduces to Ericksen-Leslie when the microinertia tensor is given by 
$j=J(\mathbb{I}-\mathbf{n}\mathbf{n}^T)$, which is a relation preserved by the dynamics; see \cite{GBRa2009}.

\subsection{Eringen's micropolar theory}

At this point, it is important to notice that, upon following the 
same reasoning as in the previous sections, one can set an initial 
condition of the type $\nabla\mathbf{n}_0=\mathbf{n}_0\times
\boldsymbol{\gamma}_0$, so that $\nabla\mathbf{n}=\mathbf{n}\times
\boldsymbol{\gamma}$ at all times, provided 
$\color{black} \boldsymbol{\widehat\gamma}= -(\nabla\chi)\chi^{-1}+
\chi{\boldsymbol{\widehat\gamma}}_0\chi^{-1}$. This 
operation would modify the Luhiller-Rey Lagrangian by inserting the 
potential  $\boldsymbol{\gamma}$ in the set of dynamical variables 
to produce a Lagrangian
\[
\ell(\boldsymbol{\nu},j,\mathbf{n},\boldsymbol{\gamma})=
\frac{1}{2}\int_\mathcal{D}\boldsymbol{\nu}\cdot j\boldsymbol{\nu}
\,\de^3 x-\int_\mathcal{D}F(\mathbf{n},\mathbf{n}\times
\boldsymbol{\gamma})\,\de^3 x\,.
\]
Now, before looking into the Euler-Poincar\'e equations for this 
Lagrangian, a crucial step consists of the observation that for
$j=J(\mathbb{I}-\mathbf{n}\mathbf{n}^T)$, the Frank free energy can 
be written as a free energy functional $\Psi(j,\bgamma)=
F(\mathbf{n},\mathbf{n}\times\boldsymbol{\gamma})=
F(\mathbf{n},\nabla\mathbf{n})$ (see \cite{GBRaTr2011} for the explicit derivation), that is {\color{black} (see Appendix \ref{appendix}),}
\begin{multline}\label{FrankMicropolar}
\Psi(j,\gamma)=\frac{K_2}{J}\operatorname{Tr}(j\boldsymbol{\gamma})+ 
\frac{K_{11}}J\Big(\!\operatorname{Tr}\!\left(
(\boldsymbol{\gamma}^A)^2\right)\left(\operatorname{Tr}(j)-
J\right)-2\operatorname{Tr}\!\left(j(\boldsymbol{\gamma}^A)^2\right)
\!\Big)
\\
+\frac12\frac{K_{22}}{J^2}\operatorname{Tr}^2(j\bgamma)
-\frac{K_{33}}J\operatorname{Tr}\!\left(\!\left((\bgamma j)^A-J\bgamma^A\right)^{\!2}\right),
\end{multline}
where the superscript $A$ denotes the antisymmetric part and the wryness tensor $\boldsymbol{\gamma}_i^a$ is considered as a $3\times 3$ real matrix, so that $\bgamma j$ is obtained by standard matrix multiplication (i.e., $(\boldsymbol{\gamma} j)_{lk}=
\boldsymbol{\gamma}_l^aj_{ak}$).
 Notice that many terms in this expression of the free energy differ from those in \cite{Eringen1993}. 
{\color{black} A particularly easy expression is available for the \emph{one-constant approximation} (i.e., $K_2=0$ and $K_{11}={K_{22}}={K_{33}}=K$):
\[
F(\bn,\nabla\bn)=\frac12K\left\|\nabla\bn\right\|^2=\frac12K\left\|\bn\times\bgamma\right\|^2=\frac{K}2\bgamma_i\cdot j\bgamma_i=\frac{K}2\operatorname{Tr}(\bgamma j\bgamma)=\Psi(j,\bgamma)
\,.
\]}
Consequently, we obtain the \emph{micropolar} Lagrangian
\[
\ell(\boldsymbol{\nu},j,\boldsymbol{\gamma})=
\frac{1}{2}\int_\mathcal{D}\boldsymbol{\nu}\cdot j\boldsymbol{\nu}
\,\de^3 x-\int_\mathcal{D}\Psi(j,\boldsymbol{\gamma})\,\de^3 x
\]
whose associated variational principle gives the Euler-Poincar\'e equations
\begin{equation}\label{EP-Mic1}
\left\{
\begin{array}{l}
\vspace{0.2cm}\displaystyle\frac{\partial}{\partial t}
\frac{\delta \ell}{\delta \boldsymbol{\nu}}=
\boldsymbol{\nu}\times \frac{\delta \ell}{\delta \boldsymbol{\nu}}
+\!\overrightarrow{\,\left[j,\frac{\delta \ell}{\delta j}\right]\,}
+\frac{\partial}{\partial x^i}
\frac{\delta\ell}{\delta\boldsymbol{\gamma}_i}
+\boldsymbol{\gamma}_i \times 
\frac{\delta \ell}{\delta \boldsymbol{\gamma}_i } \\
\vspace{0.2cm}\displaystyle\partial_t j+
[j,\widehat{\boldsymbol{\nu}}]=0\\
\displaystyle  \partial _t \boldsymbol{\gamma}_i +  
\boldsymbol{\gamma}_i \times \boldsymbol{\nu}+ 
\partial_i \boldsymbol{\nu}=0.
\end{array}\right. 
\end{equation}
A short computation, using the explicit expression of $\ell$, yields
the equations
\begin{equation}\label{EP-Mic2}
\left\{
\begin{array}{l}
\vspace{0.2cm}\displaystyle\frac{\partial}{\partial t}
j \boldsymbol{\nu}=
-\!\overrightarrow{\,\left[j,\frac{\partial \Psi }{\delta j}\right]\,}
-\frac{\partial}{\partial x^i}
\frac{\partial \Psi }{\partial \boldsymbol{\gamma}_i}
-\boldsymbol{\gamma}_i \times 
\frac{\partial \Psi}{\partial  \boldsymbol{\gamma}_i } \\
\vspace{0.2cm}\displaystyle\partial_t j+
[j,\widehat{\boldsymbol{\nu}}]=0\\
\displaystyle  \partial _t \boldsymbol{\gamma}_i +  
\boldsymbol{\gamma}_i \times \boldsymbol{\nu}+ 
\partial_i \boldsymbol{\nu}=0.
\end{array}\right. 
\end{equation}
If $ \Psi $ verifies the \emph{axiom of objectivity} (i.e., $\Psi(j,\bgamma)=\Psi(\mathcal{R}^{-1}j\mathcal{R},\mathcal{R}^{-1}\bgamma\mathcal{R}$) for any orthogonal matrix $\mathcal{R}\in O(3)$), then one can show (see \cite{GBRa2009}) that the first equation simplifies to
\[
\frac{\partial}{\partial t}
j \boldsymbol{\nu}=
-\frac{\partial}{\partial x^i}
\frac{\partial \Psi }{\partial \boldsymbol{\gamma}_i}
-\boldsymbol{\gamma}^a \times 
\frac{\partial \Psi}{\partial  \boldsymbol{\gamma}^a },
\]
where we note that the cross product is now taken relative to the manifold indices.

Notice that if the initial condition $j_0$ of $j$ is such that 
$\nabla j_0=\left[j_0,\widehat{\boldsymbol{\gamma}}_0\right]$, then 
the relation $\nabla j=\left[j,\widehat{ \boldsymbol{\gamma}}\right]$ 
is preserved by the dynamics; this can be shown by explicitly 
computing $\nabla j=\nabla(\chi j_0\chi^{-1})$.

The micropolar model derived above is an extremely powerful tool in 
ideal nematodynamics. Indeed, this model has a twofold advantage: 
first, {\color{black} it accounts for molecules of different shapes} and, second, it incorporates disclination dynamics that may 
arise from an initial potential $\boldsymbol{\gamma}_0\neq0$. Due to 
the way it is constructed, the above micropolar theory naturally 
comprises Ericksen-Leslie dynamics for uniaxial systems, which is 
recovered upon setting $\nabla j=\left[j,
 \widehat{\boldsymbol{\gamma}}\right]$ and 
$j=J(\mathbb{I}-\mathbf{n}\mathbf{n}^T)$ (both relations are 
preserved by the dynamics). Therefore, the above micropolar theory is 
the most general gauge theory of nematodynamics 
for systems composed of molecules with arbitrary shape. Indeed, this 
property follows from the fact that, if 
$j=J(\mathbb{I}-\mathbf{n}\mathbf{n}^T)$, then the micropolar free 
energy coincides with the Frank energy: $\Psi(j,\boldsymbol{\gamma})
=F(\mathbf{n},\mathbf{n}\times\boldsymbol{\gamma})=
F(\mathbf{n},\nabla\mathbf{n})$, so that the micropolar Lagrangian 
reduces to the Ericksen-Leslie Lagrangian.

{\color{black} 
\subsection{Remarks on biaxial nematic molecules}

This section explains how the microinertia tensor may be used to describe the dynamics of biaxial liquid crystals. Biaxial molecules possess two preferred directions in space, so that their order parameter is identified with a pair of orthogonal director fields. The first dynamical description of biaxial liquid crystals appears probably in \cite{VoKa1981}, where the equations of motion are derived by using a Hamiltonian approach and upon neglecting spin inertial effects. These equations were also obtained more recently in \cite{GBTr2010} by using the Euler-Poincar\'e approach and upon including spin inertial effects. The treatment follows exactly the same steps that were described in the preceding sections, upon replacing the director $\bn$ by the director pair $(\bn^{(1)},\bn^{(2)})$, so that $(\bn^{(1)},\bn^{(2)})=(\chi\bn^{(1)}_{0},\chi\bn^{(2)}_{0})$. Then, the reduction procedure outlined in Section \ref{Sec:ELReduction} yields the equations of motion \cite{GBTr2010}
\begin{equation}\label{biaxial}
\frac{\partial}{\partial t}\frac{\delta \ell}{\delta
\boldsymbol{\nu}}=\boldsymbol{\nu}\times\frac{\delta \ell}{\delta
\boldsymbol{\nu}}+{\bf n}^{(i)}\times\frac{\delta \ell}{\delta {\bf n}^{(i)}}
\,,\qquad\quad\frac{\partial {\bf n}^{(i)}}{\partial t}={\bf n}^{(i)}\times\boldsymbol\nu
\end{equation}
where the summation convention over repeated indices is used in the first equation. The question of how the Lagrangian can be expressed in terms of the two director fields (and their gradients) is still an open question in liquid crystal dynamics. As in the uniaxial case, the use of the director pair $(\bn^{(1)},\bn^{(2)})$ for describing the biaxial molecules can be replaced by the definition of an appropriate tensor order parameter. In standard textbooks (e.g., \cite{KlLa2003}, equation (11.9)), this tensor order parameter is identified with a linear combination of the following tensors
\[
\mathsf{A}=\Bbb{I}-\bn^{(1)}\bn^{(1)\,T}
,\qquad
\mathsf{B}=(\bn^{(1)}\times\bn^{(2)})(\bn^{(1)}\times\bn^{(2)})^T-\bn^{(2)}\bn^{(2)\,T}
,
\]
so that the total tensor order parameter is written as $j=\alpha\mathsf{A}+\beta\mathsf{B}$, where the constants $\alpha$ and $\beta$ are thermodynamical constants. Then, it is readily seen that, the dynamical relation $(\bn^{(1)},\bn^{(2)})=(\chi\bn^{(1)}_{0},\chi\bn^{(2)}_{0})$ implies $j=\chi j_0\chi^{-1}$, where $ j_0=\alpha\mathsf{A}_0+\beta\mathsf{B}_0$ is the initial configuration involving the reference director fields $(\bn^{(1)}_{0},\bn^{(2)}_{0})$. In turn, the relation $j=\chi j_0\chi^{-1}$ is precisely the same that appears in Eringen's theory and thus, if the Lagrangian $l(\bnu,\bn^{(1)},\bn^{(2)})$  in \eqref{biaxial} can be expressed in terms of $j$ as $\ell(\bnu,\bn^{(1)},\bn^{(2)})=l(\bnu,j)$, then the resulting equations of motion read
\begin{equation*}
\left\{
\begin{array}{l}
\vspace{0.2cm}
\displaystyle\frac{\partial}{\partial t}
\frac{\delta l}{\delta \boldsymbol{\nu}}=
\boldsymbol{\nu}\times \frac{\delta l}{\delta \boldsymbol{\nu}}
+\!\overrightarrow{\,\left[j,\frac{\delta l}{\delta j}\right]}
\\
\vspace{0.2cm}\displaystyle\frac{\partial j}{\partial t}+[j,\widehat{{\nu}}]=0
\,.
\end{array}\right. 
\end{equation*}
Notice that the Lagrangian $l$ involves a free energy term containing the gradient $\nabla j$. On the other hand, one can repeat the same steps as before, thereby introducing the wryness tensor \eqref{gamma_def} via the gauge-invariant relation $\nabla j=\left[j,\widehat{ \boldsymbol{\gamma}}\right]$. This last step  produces precisely the Euler-Poincar\'e equations \eqref{EP-Mic1}, which generalize the micropolar equations \eqref{EP-Mic2} to arbitrary Lagrangians. In this sense, equations \eqref{EP-Mic1} were hereby shown to apply to either uniaxial and biaxial molecules, depending on how the microinertia tensor is expressed in terms of the director field(s).

Notice that the above arguments require the underling assumption that there exists a Lagrangian $l$ such that $\ell(\bnu,\bn^{(1)},\bn^{(2)})=l(\bnu,j)$. This assumption may look pretty restrictive. When this assumption fails, Eringen's micropolar theory needs to be extended to Lagrangians of the type $\ell(\bnu,\bn^{(1)},\bn^{(2)})=l(\bnu,\mathsf{A},\mathsf{B})$. This step does not present any difficulty and was treated in \cite{GBTr2010}. 

Another cautionary remark is also necessary here: although Eringen's microinertia tensor may describe the dynamics of both uniaxial and biaxial nematics, the theory does not account for phase transitions, e.g., from uniaxial to biaxial. Indeed, the initial condition $\bn^{(2)}_0=0$ gives $\bn^{(2)}=\chi\bn^{(2)}_0=0$ and $\mathsf{B}=0$ at all times, so that if one starts with a uniaxial phase it is not possible to end up with a biaxial phase.
}

\section{Euler-Poincar\'e approach to flowing liquid crystals}

This section reviews briefly how all previous results can be extended 
to flowing liquid crystals. In order to deal with fluid systems, it 
is necessary to deal with Lagrangian systems on the group 
$\operatorname{Diff}(\mathcal{D})$ of diffeomorphisms in a region 
$\mathcal{D}\subset\mathbb{R}^3$, that is the Lie group of smooth 
invertible coordinate transformations on  $\mathcal{D}$. The next 
section introduces the topic by presenting the Euler-Poincar\'e 
reduction for compressible fluids.

\subsection{Euler-Poincar\'e reduction for ideal compressible fluids}

Two main descriptions are used in fluid dynamics: the Lagrangian and the Eulerian description. While the first approach 
describes the fluid flow by following an individual fluid particle, 
the second describes the flow as it is measured by a fixed observer 
in physical space. In the Lagrangian representation, the Lagrangian 
fluid coordinate is a (smooth and invertible) coordinate 
transformation $\boldsymbol\eta=\boldsymbol\eta(\mathbf{x}_0,t)$ that 
maps the initial Lagrangian label $\mathbf{x}_0$ to its position 
$\boldsymbol\eta$ at time $t$. This leads to a standard variational 
principle of the type
\[
\delta\!\int_{t_1}^{t_2}\!\mathcal{L}(\boldsymbol\eta,\dot{\boldsymbol\eta})\,\de t=0
\]
for arbitrary variations $\delta\boldsymbol{\eta}$ that vanish at 
$t_1$ and $t_2$. The choice of the Lagrangian is dictated by the 
form of kinetic and potential energy of the fluid. For example, for 
barotropic compressible fluids, the  internal (potential) energy 
depends only on the mass density, which we shall write as 
$\rho_0(\mathbf{x}_0)$ in terms of Lagrangian labels. For simplicity, 
we start with pressureless compressible fluids and we shall insert the internal energy later on. The explicit form of the kinetic 
energy part of the Lagrangian is
\[
\mathcal{L}(\boldsymbol\eta,\dot{\boldsymbol\eta})=
\frac{1}{2}\int_\mathcal{D}\!\rho_0(\mathbf{x}_0)
\left|\dot{\boldsymbol\eta}(\mathbf{x}_0,t)\right|^2\de^3\mathbf{x}_0
\,,
\]
so that the corresponding Euler-Lagrange equation reads 
$\ddot{\boldsymbol\eta}(\mathbf{x}_0,t)=0$, which is a free-motion 
equation. At this point, we notice that, contrary to the case of 
incompressible fluids (which are such that 
$\operatorname{det}(\mathcal{J}_{\boldsymbol\eta})=1$, where 
$\mathcal{J}_{\boldsymbol\eta}$ is the Jacobian matrix of 
$\boldsymbol\eta$), the relabeling symmetry is broken in the present 
situation:
\[
\mathcal{L}(\boldsymbol\eta,\dot{\boldsymbol\eta})\neq\mathcal{L}(\mathbb{I},\dot{\boldsymbol\eta}\circ\boldsymbol\eta^{-1}),
\]
where $\circ$ denotes composition of maps, so that 
$(\dot{\boldsymbol\eta}\circ\boldsymbol\eta^{-1})(\mathbf{x})=
\dot{\boldsymbol\eta}(\boldsymbol\eta^{-1}(\mathbf{x}))$. However, one can transfer the reasoning of the previous sections to the present case. Indeed, if we define the time-dependent variable
\[
\rho(\mathbf{x},t)=\int_\mathcal{D}\rho_0(\mathbf{x}_0)\,
\delta(\mathbf{x}-\boldsymbol\eta(\mathbf{x}_0,t))\,\de^3
\mathbf{x}_0=:(\boldsymbol\eta_*\rho_0)(\mathbf{x},t)
\,,
\]
one can verify that 
\[
\mathcal{L}_{\rho_0}(\boldsymbol\eta,\dot{\boldsymbol\eta})=
\mathcal{L}_{\boldsymbol\eta_*\rho_0}(\mathbb{I},
\dot{\boldsymbol\eta}\circ\boldsymbol\eta^{-1})
\,,
\]
as the following computation shows: 
\begin{align*}
\mathcal{L}_{\rho_0}(\boldsymbol\eta,\dot{\boldsymbol\eta})
=&\ \frac{1}{2}\int_{\mathcal{D}}\!\de^3 \mathbf{x}
\int_{\mathcal{D}}\de^3 \mathbf{x}_0 
\,\rho_0(\mathbf{x}_0)\,\delta(\mathbf{x}-
\boldsymbol\eta(\mathbf{x}_0,t))\left|\left(\dot{\boldsymbol\eta}
\circ\boldsymbol\eta^{-1}\right)(\mathbf{x},t)\right|^2 \\
=&\ \frac{1}{2}\int_{\mathcal{D}}\!\rho(\mathbf{x})
\left|\left(\dot{\boldsymbol\eta}\circ\boldsymbol\eta^{-1}\right)
(\mathbf{x},t)\right|^2\de^3 \mathbf{x}  \\
=&\ \mathcal{L}_{\rho}(\mathbb{I},\dot{\boldsymbol\eta}\circ
\boldsymbol\eta^{-1})
\,.
\end{align*}
The map sending $\rho_0$ to $\boldsymbol\eta_*\rho_0$ is known as 
\emph{Lagrange-to-Euler map} as it maps the Lagrangian density 
$\rho_0(\mathbf{x}_0)$ to its corresponding Eulerian quantity 
$\rho(\mathbf{x})$. In more geometric terms, one says that $\rho$ 
is given by the \emph{push-forward} of $\rho_0$ by $\boldsymbol\eta$. 
As it happened in the previous sections, the Lie group action of 
$\boldsymbol\eta$ on the parameter $\rho_0$ emerges naturally in the 
reduction process and, in this case, it is precisely given by the 
push forward $\boldsymbol\eta_*\rho_0$.

Now, if we define $\boldsymbol{u}(\mathbf{x},t):=
(\dot{\boldsymbol\eta}\circ\boldsymbol\eta^{-1})(\mathbf{x},t)$  
and $\ell(\boldsymbol{u},\rho):=\mathcal{L}_{\rho}(\mathbb{I},
\dot{\boldsymbol\eta}\circ\boldsymbol\eta^{-1})$, then we obtain 
the Euler-Poincar\'e variational principle
\[
\delta\!\int_{t_1}^{t_2}\ell(\boldsymbol{u},\rho)\,\de t=
\delta\!\int_{t_1}^{t_2}\!\int_\mathcal{D}\rho\,
\frac{\left|\boldsymbol{u}\right|^2}2\,\de^3 \mathbf{x}\ \de t=0
\]
subject to certain constrained variations of $\boldsymbol{u}$ and
$\rho$ that we now compute. {\color{black} First, we use the formula \cite{HoMaRa1998}}
\[
\delta\boldsymbol{u}=
\delta(\dot{\boldsymbol\eta}\boldsymbol\eta^{-1})=
\partial_t\left((\delta\boldsymbol\eta)\boldsymbol\eta^{-1}\right)+
\left[(\delta\boldsymbol\eta)\boldsymbol\eta^{-1},\boldsymbol{u}\right]
\]
of general validity for $\boldsymbol\eta$ belonging to an arbitrary Lie group $G$ with Lie algebra structure $(\mathfrak{g},\left[\cdot,\cdot\right])$. In the special case $G=\operatorname{Diff}(\mathcal{D})$, one has $(\delta\boldsymbol\eta)\boldsymbol\eta^{-1}=
(\delta\boldsymbol\eta)\circ\boldsymbol\eta^{-1}$ and the corresponding Lie algebra is identified with the space $\mathfrak{X}(\mathcal{D})$ of vector fields, which is endowed with the negative \emph{Jacobi-Lie bracket}
\[
\left[\boldsymbol{u},\boldsymbol{v}\right]=(\boldsymbol{v}\cdot\nabla)\boldsymbol{u}-(\boldsymbol{u}\cdot\nabla)\boldsymbol{v}\,.
\]
Next, we compute $\delta\rho$ by pairing the latter with a test function $\varphi$, as follows:
\begin{align*}
\int_\mathcal{D}\de^3 \mathbf{x}\,\delta\rho(\mathbf{x})\,
\varphi(\mathbf{x})
&=
-\int_\mathcal{D}\de^3 \mathbf{x}\,\varphi(\mathbf{x})
\int_\mathcal{D}\de^3 \mathbf{x}_0\,\rho_0(\mathbf{x}_0)
\nabla\delta(\mathbf{x}-\boldsymbol\eta(\mathbf{x}_0))\cdot
\delta{\boldsymbol\eta}(\mathbf{x}_0)
\\&=
\int_\mathcal{D}\de^3 \mathbf{x}\,\nabla\varphi(\mathbf{x})
\int_\mathcal{D}\de^3 \mathbf{x}_0\,\rho_0(\mathbf{x}_0)
\delta(\mathbf{x}-\boldsymbol\eta(\mathbf{x}_0))\cdot
(\delta{\boldsymbol\eta}\circ\boldsymbol\eta^{-1})(\mathbf{x})
\\&=
-\int_\mathcal{D}\de^3 \mathbf{x}\,\varphi(\mathbf{x})
\operatorname{div}\!\big(\rho(\mathbf{x})(\delta{\boldsymbol\eta}\circ\boldsymbol\eta^{-1})(\mathbf{x})\big)
\,,
\end{align*}
so that
\[
\delta\rho=-\operatorname{div}\!\big(\rho(\delta{\boldsymbol\eta}\circ\boldsymbol\eta^{-1})\big).
\]
In conclusion, the Euler-Poincar\'e equations are
\begin{equation}\label{EP_general}
\left\{
\begin{array}{l}
\vspace{0.2cm}\displaystyle
\frac{\partial}{\partial t}\frac{\delta \ell}{\delta \boldsymbol{u}}
+(\boldsymbol{u}\cdot\nabla)\frac{\delta \ell}{\delta \boldsymbol{u}}
+\operatorname{div}(\boldsymbol{u})\frac{\delta \ell}{\delta 
\boldsymbol{u}}+\nabla\boldsymbol{u}^T\cdot\frac{\delta \ell}{\delta 
\boldsymbol{u}}=\rho\nabla\frac{\delta \ell}{\delta \rho}
\\
\displaystyle\frac{\partial\rho}{\partial t}+\operatorname{div}\!\big(\rho
\boldsymbol{u}\big)=0
\end{array}
\right.
\end{equation} 
that is, after computing $\delta\ell/\delta\boldsymbol{u}=
\rho\boldsymbol{u}$ and $\delta\ell/\delta\rho=\left|\boldsymbol{u}
\right|^2\!/2$ and some calculations,
\[
\left\{
\begin{array}{l}
\vspace{0.2cm}\displaystyle
\frac{\partial\boldsymbol{u}}{\partial t}+(\boldsymbol{u}\cdot
\nabla)\boldsymbol{u}=0   \\
\displaystyle\frac{\partial\rho}{\partial t}+\operatorname{div}\!
\big(\rho\boldsymbol{u}\big)=0
\end{array}
\right .
\]
which are the equations for a pressureless compressible fluid with 
\emph{Eulerian velocity} $\boldsymbol{u}$. If we were considering 
(barotropic) pressure effects, it would suffice to add the internal 
energy term in the new Lagrangian
\[
\ell(\boldsymbol{u},\rho)=\int_\mathcal{D}\rho\,\frac{\left|
\boldsymbol{u}\right|^2}2\,\de^3 \mathbf{x}-
\int_\mathcal{D}\rho\,\mathcal{U}
(\rho)\,\de^3\mathbf{x}
\,.
\]
Then, the Euler-Poincar\'e equations for the velocity would be 
modified by the inclusion of a pressure term as follows:
\[
\left\{
\begin{array}{l}
\vspace{0.2cm}\displaystyle\frac{\partial\boldsymbol{u}}{\partial t}+(\boldsymbol{u}\cdot
\nabla)\boldsymbol{u}=-\frac{1}{\rho } \nabla\mathsf{p} \\
\displaystyle\frac{\partial\rho}{\partial t}+\operatorname{div}\!\big(\rho
\boldsymbol{u}\big)=0,
\end{array}
\right .
\]
where the pressure is given by $\mathsf{p}=\rho^2\mathcal{U}'(\rho)$. 
Notice that the unreduced Lagrangian $\mathcal{L}_{\rho_0}
(\boldsymbol\eta,\dot{\boldsymbol\eta})$ for barotropic compressible 
fluids can be reconstructed from $\ell(\boldsymbol{u},\rho)$ by 
simply replacing $\boldsymbol{u}=\dot{\boldsymbol\eta}\circ
\boldsymbol\eta^{-1}$ and $\rho=\boldsymbol\eta_*\rho_0$.

At this point, one may wonder what the correspondent of equation \eqref{Noether-quantity} is in this case. In the context of fluids, the operation  $\color{black} \operatorname{Ad}^*_{\chi}\delta\ell/\delta\nu$ corresponds to the {\color{black} pull back} $\boldsymbol\eta^*(\delta\ell/\delta\boldsymbol{u})$ and the main ingredient is the dual of the inclusion $i:\mathfrak{X}_{\rho_0}(\mathcal{D})\hookrightarrow\mathfrak{X}(\mathcal{D})$. Here, $\mathfrak{X}_{\rho_0}(\mathcal{D})$ denotes the Lie algebra of vector fields preserving $\rho_0(\bx_0)\,\de^3\bx_0$, so that $\boldsymbol\xi\in\mathfrak{X}_{\rho_0}(\mathcal{D}) \Leftrightarrow\operatorname{div}(\rho_0\,\boldsymbol\xi)=0$ and $i(\boldsymbol\xi)=\boldsymbol\xi$. Consequently, $\boldsymbol\xi\in\mathfrak{X}_{\rho_0}(\mathcal{D})$ if and only if there exists a vector potential $\boldsymbol\psi$ such that $\boldsymbol\xi=\rho_0^{-1}\operatorname{curl}\boldsymbol\psi$, so that the Lie algebra $\mathfrak{X}_{\rho_0}(\mathcal{D})$ can be identified with the Lie algebra of vector potentials. Therefore, upon taking the $L^2$-pairing of the latter relation with $\boldsymbol\eta^*(\delta\ell/\delta\boldsymbol{u})$, one computes
\[\color{black} 
i^*\!\left(\boldsymbol\eta^*\frac{\delta\ell}{\delta\boldsymbol{u}}\right)=\operatorname{curl}\!\left(\frac1{\rho_0}\, \boldsymbol\eta^*\frac{\delta\ell}{\delta\boldsymbol{u}}\right)=\operatorname{curl}\!\left(\frac1{\rho_0(\bx_0)}\int_\mathcal{D}\frac{\delta\ell}{\delta\boldsymbol{u}}(\bx,t)\,\delta\big(\bx_0-\boldsymbol\eta^{-1}(\bx,t)\big)\,\de^3\bx\right)
\]
Then, upon considering a closed surface in $\mathcal{D}$ with boundary $\gamma_0$, one can easily apply Stokes' theorem to write
\[
0=\frac{\de}{\de t}\oint_{\gamma_0}\frac1{\rho_0}\, \boldsymbol\eta^*\frac{\delta\ell}{\delta\boldsymbol{u}}\cdot\de\bx_0=\frac{\de}{\de t}\oint_{\boldsymbol\eta(\gamma_0)}\frac1{\rho}\, \frac{\delta\ell}{\delta\boldsymbol{u}}\cdot\de\bx=\frac{\de}{\de t}\oint_{\gamma\!}\boldsymbol{u}\cdot\de\bx
\,,
\]
where $\gamma=\boldsymbol\eta(\gamma_0)$ and the second step follows by 
changing coordinates {\color{black} (using the relation $\rho=\boldsymbol\eta_*\rho_0$)}. 
The above conservation relation is of paramount importance in fluid 
mechanics and it is known under the name of \emph{Kelvin-Noether 
theorem}.

\begin{remark}{\rm 
Notice that one can also use the $L^2$-paring to identify the dual spaces 
as $\mathfrak{X}(\mathcal{D})^*= \mathfrak{X}(\mathcal{D})$ and 
$\mathfrak{X}_{\rho_0}(\mathcal{D})^*= 
\mathfrak{X}_{\frac{1}{\rho_0}}(\mathcal{D})$. Then we have  
$i^\ast (\boldsymbol{u})= \rho_{0\,} P \!\left(\rho_0^{-1}\boldsymbol{u} 
\right) $, where $P$ denotes the Hodge projection onto divergence free 
vector fields. A direct computation using \eqref{EP_general} then shows 
that
\[
\frac{d}{dt}\!\left( i^*\! \left( \boldsymbol\eta ^* 
\frac{\delta \ell}{\delta \boldsymbol{u} } \right) \!\right)=0.
\]
This corresponds to Noether's theorem associated to the isotropy subgroup 
of $ \rho _0 $, {\color{black} as discussed in \S\ref{cons_quant}.}}
\end{remark}

\subsection{Euler-Poincar\'e reduction by stages for liquid crystals}

This section extends the previous discussion on compressible fluids 
to consider nematic liquid crystals. As we shall see, this reduction 
proceeds by two stages.

For flowing liquid crystals, the simplest form of Hamilton's 
principle is
\[
\delta\int^{t_1}_{t_0}\!L_{(\bn_0,\rho_0)}(\boldsymbol\eta,
\dot{\boldsymbol\eta},\chi,\dot{\chi})\,\de t=0,
\]
where $\boldsymbol\eta$ determines a Lagrangian trajectory on the Lie 
group of diffeomorphisms while $\chi(\mathbf{x}_0)\in SO(3)$ 
identifies the rotational state of the point $\mathbf{x}_0$. The 
initial director field $\mathbf{n}_0(\mathbf{x}_0)$ and mass 
density $\rho_0(\mathbf{x}_0)$ are just time-independent 
parameters appearing in the expression of the Lagrangian. By 
extension of the Ericksen-Leslie Lagrangian, we write
\begin{multline}
\mathcal{L}_{(\mathbf{n}_0,\rho_0)}(\boldsymbol\eta,
\dot{\boldsymbol\eta},\chi,\dot\chi)=
\frac{1}{2}\int_\mathcal{D}\! \rho_0(\mathbf{x}_0)
\left|\dot{\boldsymbol\eta}(\mathbf{x}_0)\right|^2\de^3\mathbf{x}_0+
\frac{J}{2}\int_\mathcal{D}\! \rho_0(\mathbf{x}_0)
\left|\dot{\chi}(\mathbf{x}_0)\mathbf{n}_0(\mathbf{x}_0)\right|^2 
\de^3\mathbf{x}_0    \\
-\int_\mathcal{D}\! \rho_0(\mathbf{x}_0)\,F\!\left(
(\boldsymbol\eta_*\rho_0)\circ\boldsymbol\eta,(\chi\mathbf{n}_0)
\circ\boldsymbol\eta^{-1},\!\nabla\big((\chi\mathbf{n}_0)\circ
\boldsymbol\eta^{-1}\right)\de^3\mathbf{x}_0\,.
\end{multline}
In the first stage reduction, we introduce the variables 
$\widehat{\boldsymbol{\omega}}(\mathbf{x}_0)=
\dot\chi(\mathbf{x}_0)\chi^{-1}(\mathbf{x}_0)$ and 
$\bar{\mathbf{n}}(\mathbf{x}_0)=\chi(\mathbf{x}_0)\mathbf{n}_0(\mathbf{x}_0)$. Then, the Lagrangian possesses the following symmetry
\[
\mathcal{L}_{(\mathbf{n}_0,\rho_0)}(\boldsymbol\eta,
\dot{\boldsymbol\eta},\chi,\dot\chi)=
\mathcal{L}_{(\bar{\mathbf{n}},\rho_0)}(\boldsymbol\eta,
\dot{\boldsymbol\eta},\chi\chi^{-1},\dot\chi\chi^{-1})=:
\bar{\ell}_{\rho_0}(\boldsymbol\eta,\dot{\boldsymbol\eta},
\boldsymbol\omega,\bar{\mathbf{n}}).
\]
However, this second Lagrangian also possesses another type of symmetry that is related to the relabeling properties. Upon defining the variables 
\[
\boldsymbol{\nu}:=\boldsymbol\omega\circ\boldsymbol\eta^{-1}\,,
\qquad\mathbf{n}:=\bar{\mathbf{n}}\circ\boldsymbol\eta^{-1}
\,,
\]
this symmetry is
\[
\bar{\ell}_{\rho_0}(\boldsymbol\eta,\dot{\boldsymbol\eta},
\boldsymbol\omega,\bar{\mathbf{n}})=
\bar{\ell}_{\boldsymbol\eta_*\rho_0}(\boldsymbol\eta\circ
\boldsymbol\eta^{-1},\dot{\boldsymbol\eta}\circ\boldsymbol\eta^{-1},
\boldsymbol\omega\circ\boldsymbol\eta^{-1},\bar{\mathbf{n}}\circ
\boldsymbol\eta^{-1})=:\ell(\boldsymbol{u},\rho,\boldsymbol{\nu},
\mathbf{n}),
\]
where the action of the diffeomorphism group again emerges in the definition of the variables $\boldsymbol{\nu}$ and $\mathbf{n}$ and
\[
\ell(\boldsymbol{u},\rho,\boldsymbol{\nu},\mathbf{n})=
\frac{1}{2}\int_\mathcal{D}\! \rho\left|\boldsymbol{u}\right|^2\de^3 
\mathbf{x}+
\frac{J}{2}\int_\mathcal{D}\! \rho\left|\boldsymbol{\nu}\times
\mathbf{n}\right|^2\de^3 \mathbf{x}-
\int_\mathcal{D}\! \rho\,F(\rho^{-1} ,\mathbf{n},\nabla\mathbf{n})\,\de^3 \mathbf{x}\,.
\]
At this point, upon denoting $\boldsymbol{w}=\delta\boldsymbol{\eta}\circ\boldsymbol{\eta}^{-1}$ and $\boldsymbol\xi=(\delta\chi)\chi^{-1}\circ\boldsymbol{\eta}^{-1}$, the variational principle
\[
\delta\int_{t_1}^{t_2}\ell(\boldsymbol{u},\rho,\boldsymbol{\nu},\mathbf{n})\,\de t=0
\]
for the constrained variations (see \cite{GBRa2009} for the derivation of these formulas) 
\begin{align*}
&\delta\boldsymbol{u}=
\partial_t\boldsymbol{w}+
\left[\boldsymbol{w},
\boldsymbol{u}\right]
\\
&\delta\rho=-\operatorname{div}\!\big(\rho{\boldsymbol{w}}
\big)
\\
&\delta \boldsymbol{\nu}=\partial_t\boldsymbol\xi-\boldsymbol{w}\cdot\nabla\bnu+\boldsymbol{u}\cdot\nabla\boldsymbol\xi-\boldsymbol\nu\times\boldsymbol\xi
\\
&\delta\mathbf{n}=-\boldsymbol{w}\cdot\nabla\bn+\boldsymbol\xi\times\bn
\,
\end{align*} 
yields the Euler-Poincar\'e equations
\begin{equation}\label{EL-LiqXal}
\left \{
\begin{array}{l}
\vspace{0.2cm}\displaystyle\frac{\partial}{\partial t}\frac{\delta \ell}{\delta \boldsymbol{u}}
+(\boldsymbol{u}\cdot\nabla)\frac{\delta \ell}{\delta \boldsymbol{u}}
+\operatorname{div}(\boldsymbol{u})
\frac{\delta \ell}{\delta \boldsymbol{u}}+
\nabla\boldsymbol{u}^T\cdot\frac{\delta \ell}{\delta \boldsymbol{u}}
=-\nabla\boldsymbol{\nu}\cdot\frac{\delta l}{\delta\boldsymbol{\nu}}-
\nabla\mathbf{n}^T\cdot\frac{\delta l}{\delta \mathbf{n}}+
\rho\nabla\frac{\delta l}{\delta \rho}   \\
\vspace{0.2cm}\displaystyle\frac{\partial}{\partial t}\frac{\delta l}{\delta\boldsymbol{\nu}}+
\partial_i\!\left(\!u^i\frac{\delta l}{\delta\boldsymbol{\nu}}\right)
=\boldsymbol{\nu}\times\frac{\delta l}{\delta\boldsymbol{\nu}}+
\mathbf{n}\times\frac{\delta l}{\delta\mathbf{n}}  \\
\displaystyle\frac{\partial\rho}{\partial t}+\partial_i(u^i\rho)=0\,,
\hspace{1.5cm}
\frac{\partial\mathbf{n}}{\partial t}+
(\boldsymbol{u}\cdot\nabla)\mathbf{n}=\boldsymbol{\nu}\times
\mathbf{n}\,.
\end{array}
\right.
\end{equation} 
Taking the variational derivatives of the Lagrangian $\ell$ given 
above yields the Ericksen-Leslie equations of nematic liquid 
crystals
\[
\left \{
\begin{array}{l}
\vspace{0.2cm}\displaystyle\rho\left(\frac{\partial\boldsymbol{u}}{\partial t}+
(\boldsymbol{u}\cdot\nabla)\boldsymbol{u}\right)=
-\nabla\mathsf{p}
-\partial_i\!\left(\!\rho\,\nabla\mathbf{n}\cdot
\frac{\partial F}{\partial{\mathbf{n},_i}}\right)\\
\displaystyle J\frac{D^2\mathbf{n}}{Dt^2}-\left(\mathbf{n\cdot h}+
J\mathbf{n}\cdot\frac{D^2\mathbf{n} }{Dt^2}\right)\mathbf{n}+
\mathbf{h}=0\,,
\end{array}
\right.
\]
where we have defined the pressure 
$\mathsf{p}:=-\partial F/\partial\rho^{-1}$ and used the material 
derivative notation $D/Dt:=\partial_t+\boldsymbol{u}\cdot\nabla$.

At this point, one can repeat all the steps that were followed in 
the flowless case and insert an initial condition 
$\nabla\mathbf{n}_0=\mathbf{n}_0\times\boldsymbol{\gamma}$. Then, 
the relation $\nabla\mathbf{n}=\mathbf{n}\times\boldsymbol{\gamma}$ 
is again invariant, provided the diffeomorphism $\boldsymbol\eta$ 
acts on it by push-forward. An analogous result also holds for the introduction of the microinertia tensor $j$. At the end of the various steps of the reduction, one obtains a Lagrangian 
$\textcolor{black}{ \ell(\boldsymbol{u}, \boldsymbol{\nu}, \rho ,j,\boldsymbol{\gamma})}$ that reads
\[
\textcolor{black}{\ell(\boldsymbol{u}, \boldsymbol{\nu},\rho , j,\boldsymbol{\gamma})}=
\frac{1}{2}\int_\mathcal{D}\! \rho\left|\boldsymbol{u}\right|^2\de^3 \mathbf{x}+
\frac{1}{2}\int_\mathcal{D} \textcolor{black}{ \rho }\, \boldsymbol{\nu}\cdot j\boldsymbol{\nu}\,\de^3 \mathbf{x}-
\int_\mathcal{D}\textcolor{black}{ \rho \Psi(\rho ^{-1} , j,\boldsymbol{\gamma})}\,\de^3 \mathbf{x}\,,
\]
where the free energy is such that $\textcolor{black}{ \Psi(\rho ^{-1} , j,\boldsymbol{\gamma})=
F(\rho ^{-1} , \mathbf{n},\nabla\mathbf{n})}$ when $j=J(\mathbb{I}-
\mathbf{n}\mathbf{n}^T)$. In turn, upon making use of the index summation convention, this produces the system
\begin{equation}\label{EP-MicF}
\left\{
\begin{array}{l}
\vspace{0.2cm}
\displaystyle \textcolor{black}{ \frac{\partial }{\partial t}\frac{\delta \ell}{\delta \boldsymbol{u}}
+(\boldsymbol{u}\cdot\nabla)\frac{\delta \ell}{\delta \boldsymbol{u}}
+\operatorname{div}(\boldsymbol{u})
\frac{\delta \ell}{\delta \boldsymbol{u}}+
\nabla\boldsymbol{u}^T\cdot\frac{\delta \ell}{\delta \boldsymbol{u}}}\\
\vspace{0.2cm}
\displaystyle
\qquad\qquad\qquad =-\nabla\boldsymbol{\nu}\cdot\frac{\delta l}{\delta\boldsymbol{\nu}} +
\rho\nabla\frac{\delta l}{\delta \rho}  -  \frac{\delta \ell}{\delta j_{hk}}\nabla j_{hk}-\frac{\delta \ell}{\delta {\gamma}^a_k }\nabla\gamma^a_k+  \frac{\partial}{\partial x^k}\! \left(\frac{\delta \ell}{\delta \gamma^a_k}\, \bgamma^a\right)  \\
\vspace{0.2cm}\displaystyle\frac{\partial}{\partial t}
\frac{\delta \ell}{\delta \boldsymbol{\nu}}+
\partial_i\!\left(u^i\,\frac{\delta \ell}{\delta \boldsymbol{\nu}}\right)
=\boldsymbol{\nu}\times \frac{\delta \ell}{\delta \boldsymbol{\nu}}+\textcolor{black}{ 
\!\overrightarrow{\,\left[j,\frac{\delta \ell }{\delta j} \right]\,}}+
\frac{\partial}{\partial x^i}
\frac{\delta\ell}{\delta\boldsymbol{\gamma}_i}+ 
\boldsymbol{\gamma}_i \times 
\frac{\delta \ell}{\delta \boldsymbol{\gamma}_i} \\
\vspace{0.2cm}
\displaystyle\displaystyle\textcolor{black}{\frac{\partial\rho}{\partial t}+\partial_i(u^i\rho)=0},\hspace{1.5cm}\partial_t j+[j,\widehat{\boldsymbol{\nu}}]=0,\hspace{1.5cm}\partial_t \bgamma_i + \boldsymbol{\gamma}_i \times 
\boldsymbol{\nu}+ \partial_i \boldsymbol{\nu}=0
\,,
\end{array}\right. 
\end{equation}
whose first two equations become
\textcolor{black}{ \begin{equation}\label{micropolar_L_C}
\left\lbrace
\begin{array}{ll}
\vspace{0.2cm}\displaystyle\rho\left(\frac{\partial\boldsymbol{u}}{\partial t}+(\boldsymbol{u}\cdot
\nabla)\boldsymbol{u}\right)=\nabla \frac{\partial
\Psi}{\partial\rho^{-1}}-\partial_k\left(\rho\frac{\partial\Psi}{\partial\boldsymbol{\gamma}^a_k}\boldsymbol{\gamma}^a\right),\\
\displaystyle j\frac{D}{d
t}\boldsymbol{\nu}-(j\boldsymbol{\nu})\times\boldsymbol{\nu}=-\frac{1}{\rho}\operatorname{div}\left(\rho\frac{\partial\Psi}{\partial\boldsymbol{\gamma}}\right)+\boldsymbol{\gamma}^a\times\frac{\partial\Psi}{\partial\boldsymbol{\gamma}^a},
\end{array} \right.
\end{equation}
when $ \Psi $ verifies the axiom of objectivity, see \cite{GBRa2009} for details.}

{\color{black} 
Notice that, although here we presented the equations for compressible flows, their incompressible version is easily derived by setting $
\rho=1$ and by adding the pressure force $-\nabla\mathsf{p}$ on the right hand side of the velocity equation, which then becomes
\[
\frac{\partial\boldsymbol{u}}{\partial t}+(\boldsymbol{u}\cdot
\nabla)\boldsymbol{u}=-\nabla\mathsf{p}-\partial_k\left(\frac{\partial\Psi}{\partial\boldsymbol{\gamma}^a_k}\boldsymbol{\gamma}^a\right)
.
\]
The last step, i.e., adding the pressure force term, is quite common in fluid mechanics and it amounts to projecting the whole equation to its divergence-less part. Alternatively, the pressure $\mathsf{p}$ can be inserted as a Lagrange multiplier in Hamilton's principle to constrain the relation $\rho=1$; see \cite{HoMaRa1998} for explicit calculations.
}

\subsection{Conserved quantities and helicity of uniaxial nematics}

Conserved quantities also appear for nematic liquid crystals, at least in the case of uniaxial molecules. For example, it is a simple verification to see that equations \eqref{EL-LiqXal} produce the following dynamics of the momentum map $\bn\cdot\delta\ell/\delta\bnu$:
\[
\frac{\partial}{\partial t}\left(\frac{\delta\ell}{\delta\bnu}\cdot\bn\right)+\operatorname{div}\!\left(\!\boldsymbol{u}\!\left(\frac{\delta\ell}{\delta\bnu}\cdot\bn\right)\!\right)=0
\]
so that the corresponding integral is constant.

On the other hand, one may ask whether more general dynamical invariant{\color{black}s} may appear in liquid crystal dynamics. For example, it would be natural to wonder if the celebrated helicity invariant of Euler's fluid equation may have a correspondent in liquid crystals. A positive answer was given in \cite{GBTr2010,GBTr2011}, where it was shown that
\[
\frac{\de}{\de t}\oint_{\gamma}\frac1\rho\left(\frac{\delta\ell}{\delta\boldsymbol{u}}-\frac{\delta\ell}{\delta\bnu}\cdot(\nabla\bn\times\bn)\right)\cdot\de\bx
=
\frac{\de}{\de t}\oint_{\gamma}\!\big(\boldsymbol{u}-J\bnu\cdot(\nabla\bn\times\bn)\big)\cdot\de\bx=0
\]
for any loop $\gamma$ moving with the fluid velocity $\boldsymbol{u}$. Such a Kelvin-Noether circulation theorem is produced by the fluid relabeling symmetry underlying liquid crystal dynamics \cite{HoMaRa1998}. Fluid models do not always possess such conservation laws, so that the right hand side is often expressed as a non-zero circulation integral (this is the case, for example, of magnetohydrodynamics in plasma physics). However, the case of liquid crystals benefits from a general theorem in geometric mechanics that is due to \cite{KrMa1987} (see Proposition 2.2 therein). This result is based on the momentum map $-(\nabla\bn\times\bn)\cdot{\delta\ell/\delta\bnu}$ that is produced by the action of the diffeomorphisms on the Poisson manifold $\mathfrak{so}(3)^*\times S^2$, with coordinates $(\boldsymbol\sigma,\bn)$. Then, the total circulation quantity $\delta\ell/\delta\boldsymbol{u}-(\nabla\bn\times\bn)\cdot{\delta\ell/\delta\bnu}$ identifies the momentum map emerging from the diffeomorphism action on $\mathfrak{X}(\mathcal{D})^*\times\mathfrak{so}(3)^*\times S^2$, which in turn generates the Kelvin-Noether theorem. {\color{black} See \cite{GBTr2010,GBTr2011} for further details.} At this point, taking the curl of the total momentum map gives the expression of the hydrodynamic vorticity
\[
\boldsymbol\Omega=\operatorname{curl}\!\big(\boldsymbol{u}-J\bnu\cdot(\nabla\bn\times\bn)\big)
\]
and one verifies that the hydrodynamic helicity is constant
\[
\mathcal{H}=\int_\mathcal{D}\big(\boldsymbol{u}-J\bnu\cdot(\nabla\bn\times\bn)\big)\cdot\operatorname{curl}\!\big(\boldsymbol{u}-J\bnu\cdot(\nabla\bn\times\bn)\big)\,\de^3\bx
\,.
\]
The latter relation does not come as a surprise in geometric mechanics. Indeed, it is simply given by the usual expression of fluid helicity upon shifting the fluid momentum by the quantity $-J\rho\bnu\cdot(\nabla\bn\times\bn)$. {\color{black} In turn,  as noted earlier, the latter emerges from a momentum map; this type of shift by a momentum map agrees with} the general theory developed in \cite{KrMa1987} (see Corollary 2.3 therein).
It is not the purpose of this paper to present how this more involved theory works; however, it is still worth emphasizing that the geometric setting underlying nematodynamics allows to find conserved quantities and circulation theorems by simple application of general principles in geometric mechanics.

{\color{black} 
\section{Remarks on dissipative dynamics}

The dissipative case can be easily handled by using a Rayleigh dissipation function $R_{ \mathbf{n}_0 }( \chi , \dot \chi )$, i.e., 
$\left\langle \frac{\partial R_{ \mathbf{n} _0 }}{\partial \dot \chi } , \dot \chi  \right\rangle \geq 0$ (see, e.g., 
\cite[\S7.8]{MaRa2004}), having the same invariance properties as the Lagrangian $\mathcal{L}_{\bn_0}(\chi,\dot\chi)$ and thus producing a reduced function $r(\boldsymbol{\nu} , \mathbf{n} )= r( \dot \chi \chi ^{-1} , \chi \mathbf{n} _0 )=R_{ \mathbf{n}_0 }( \chi , \dot \chi )$. The associated variational principle
\[
\delta\int_{t_1}^{t_2}\mathcal{L} _{ \mathbf{n} _0 }( \chi , \dot \chi )\,\de t=\int_{t _1 } ^{ t _2 } \frac{\partial R_{ \mathbf{n} _0 }}{\partial \dot \chi } ( \chi , \dot \chi ) \cdot \delta \chi .
\,,
\]
yields, by reduction, the variational principle
\begin{equation}\label{RedHamPrinc1}
\delta\int_{t_1}^{t_2}\!\ell(\bnu,\bn)\, \de t=\int_{t_1}^{t_2}\!\frac{\delta r}{\delta\boldsymbol{\nu}} (\bnu,\bn)\cdot \delta \chi \chi ^{-1} \, \de t,
\end{equation}
for variations $\delta\mathbf{n}=
(\delta\chi)\chi^{-1}\mathbf{n}$ and $\delta\widehat{ \boldsymbol{\nu}}=\partial_t\left((\delta\chi)\chi^{-1}\right)+
\left[(\delta\chi)\chi^{-1},\widehat{\boldsymbol{\nu}}\right]$.
We thus obtain the reduced Euler-Poincar\'e equations with dissipation
(these are the reduced Lagrange-d'Alembert equations when the external force is dissipative and the configuration space is a Lie group)
\begin{equation}\label{EP-eqns1}
\left\{
\begin{array}{l}
\vspace{0.2cm}\displaystyle\frac{\partial}{\partial t}\frac{\delta \ell}{\delta \boldsymbol{\nu}}+ \frac{\delta \ell}{\delta \boldsymbol{\nu}}\times \boldsymbol{\nu}=\mathbf{n} \times \frac{\delta \ell }{\delta \mathbf{n} }- \frac{\delta  r}{\delta  \boldsymbol{\nu} }( \boldsymbol{\nu} , \mathbf{n} ) \\
\displaystyle \frac{\partial \bn}{\partial t} + \mathbf{n} \times \boldsymbol{\nu }=0.
\end{array}\right.
\end{equation}

Dissipation in Eringen micropolar theory can be handled similarly, by considering a Rayleigh dissipation function $R_{( j_0, \boldsymbol{\gamma} _0 ) }( \chi , \dot \chi )$, invariant under the isotropy subgroup of $( j_0, \boldsymbol{\gamma} _0 )$, thus yielding a function
\[
r(\boldsymbol{\nu} ,j,\boldsymbol{\gamma})= r\left(  \dot \chi \chi ^{-1} , \chi j_0  \chi  ^{-1} ,-(\nabla\chi)\chi^{-1}+\chi{\boldsymbol{\widehat\gamma}}_0\chi^{-1} \right) =R_{( j_0, \boldsymbol{\gamma} _0 ) }( \chi , \dot \chi ).
\]
Applying the same variational principle as above, we get the following modification of equation \eqref{EP-Mic1}:
\[
\left\{
\begin{array}{l}
\vspace{0.2cm}\displaystyle\frac{\partial}{\partial t}
\frac{\delta \ell}{\delta \boldsymbol{\nu}}=
\boldsymbol{\nu}\times \frac{\delta \ell}{\delta \boldsymbol{\nu}}
+\!\overrightarrow{\,\left[j,\frac{\delta \ell}{\delta j}\right]\,}
+\frac{\partial}{\partial x^i}
\frac{\delta\ell}{\delta\boldsymbol{\gamma}_i}
+\boldsymbol{\gamma}_i \times 
\frac{\delta \ell}{\delta \boldsymbol{\gamma}_i }- \frac{\delta  r}{\delta  \boldsymbol{\nu} }( \boldsymbol{\nu} , j, \boldsymbol{\gamma} ) \\
\vspace{0.2cm}\displaystyle\partial_t j+
[j,\widehat{\boldsymbol{\nu}}]=0\\
\displaystyle  \partial _t \boldsymbol{\gamma}_i +  
\boldsymbol{\gamma}_i \times \boldsymbol{\nu}+ 
\partial_i \boldsymbol{\nu}=0.
\end{array}\right. 
\]
The case of flowing liquid crystals can be handled similarly. The balance of momentum and the balance of moment of momentum equations in \eqref{EP-MicF} are modified by the addition of the terms
\[
- \frac{\delta  r}{\delta \mathbf{u}  }( \mathbf{u} ,\boldsymbol{\nu} ,\rho , j, \gamma )\quad \text{and} \quad - \frac{\delta  r}{\delta \boldsymbol{\nu}  }( \mathbf{u} ,\boldsymbol{\nu} ,\rho , j, \gamma ), 
\]
respectively.
\begin{remark}[Dissipation function]{\rm
Notice that the above equations involve an arbitrary dissipation function $r$. For example, this is the approach followed in fundamental treatments of liquid crystal dynamics, e.g., \cite{Volovick1980}. However, it would be interesting to specialize the above relations to a particular expression of $r$ and study how one can pass from one representation to the other, as it was done for $F( \mathbf{n} , \nabla \mathbf{n} )$ and $ \Psi (j, \boldsymbol{\gamma} )$ in Appendix \ref{appendix}. We leave this interesting question as an open subject for future work.}
\end{remark}
}

\medskip

\paragraph{Acknowledgments.} Stimulating conversations with David Chillingworth,  Giovanni De Matteis, and Giuseppe Gaeta are greatly acknowledged. Also, the authors wish to warmly thank the referees for their valuable comments and keen remarks that helped improving this paper.

\medskip

\appendix

{\color{black} 
\section{The free energy}\label{appendix} 

We reproduce here the computation in \cite{GBRaTr2011} that shows
how all terms in the Frank energy
 \begin{align*}
F(\mathbf{n},{\nabla}\mathbf{n})&:=K_2\underbrace{(\mathbf{n}\cdot\operatorname{curl}\mathbf{n})}_{\textbf{chirality}}+\frac{1}{2}K_{11}\underbrace{(\operatorname{div}\mathbf{n})^2}_{\textbf{splay}}+\frac{1}{2}K_{22}\underbrace{(\mathbf{n}\cdot\operatorname{curl}\mathbf{n})^2}_{\textbf{twist}}
+\frac{1}{2}K_{33}\underbrace{\|\mathbf{n}\times\operatorname{curl}\mathbf{n}\|^2}_{\textbf{bend}},
\end{align*}
can be rewritten in terms of the variables $j=J(\mathbf{I}-\bn\otimes\bn)$ and $\bgamma$, the latter being introduced through the invariant relation $\nabla\bn=\bn\times\bgamma$. Thus, the explicit expression for the micropolar free energy $\Psi(j,\gamma)$ of nematic media given in \eqref{FrankMicropolar} will be written after computing the micropolar expression for each term in the Frank energy. Further details are given in \cite{GBRaTr2011}.

\paragraph{Twist.} Using $\mathbf{n}\otimes\mathbf{n}=\mathbf{I}-{j}/J$, we have
\[
\mathbf{n}\cdot \nabla\times\mathbf{n}=-\mathbf{n}\cdot\boldsymbol{\gamma}(\mathbf{n})+\|\mathbf{n}\|^2\operatorname{Tr}(\boldsymbol{\gamma})=\frac{1}{J}\operatorname{Tr}\left({j}\bgamma\right)=\frac{1}{J}\operatorname{Tr}\left({j}\bgamma^S\right).
\]
where $\boldsymbol{\gamma}(\mathbf{n})=\bgamma_{ia}n_a$, with $a$ being the $\mathfrak{so}(3)\simeq\Bbb{R}^3$-index, and $\bgamma^S$ denotes the skew part of $\bgamma$, i.e., $\bgamma^S=\left(\bgamma-\bgamma^T\right)\!/2$, where we see $\bgamma$ as a $3\times 3$ matrix with components $\bgamma_{ia}$.

\paragraph{Splay.} We introduce the vector ${\vec{\bgamma}\!}_b=\epsilon_{abc}\bgamma_{ac}$, defined by the condition $\vec{\boldsymbol{\gamma }} \cdot\mathbf{u}=\operatorname{Tr}(\mathbf{u}\times\boldsymbol{\gamma})$, for all $\mathbf{u}\in\mathbb{R}^3$, where $\mathbf{u}\times\boldsymbol{\gamma}$ is the matrix with components $(\mathbf{u}\times\boldsymbol{\gamma})_{ia}=(\mathbf{u}\times\boldsymbol{\gamma}_i)_a$. We compute
\begin{align*}
\left(\operatorname{div}\mathbf{n}\right)^2&=( \vec{\boldsymbol{\gamma }} \cdot \mathbf{n} )(\vec{\boldsymbol{\gamma }} \cdot \mathbf{n})=\vec{\boldsymbol{\gamma }} \cdot ( \mathbf{n} \otimes \mathbf{n} ) \vec{\boldsymbol{\gamma }}\\
&=\vec{\boldsymbol{\gamma }} \cdot ( \mathbf{I} - {j}/J) \vec{\boldsymbol{\gamma }}= \|\vec{\boldsymbol{\gamma }}\|^2- \frac{1}{J} \vec{\boldsymbol{\gamma }}\cdot{j}\vec{\boldsymbol{\gamma }}\\
&=2\left(\operatorname{Tr}(j)/J-1\right)\operatorname{Tr}\left((\bgamma^A)^2\right)-\frac{4}{J}\operatorname{Tr}\left(j(\bgamma^A)^2\right),
\end{align*} 
where $\bgamma^A$ denotes the skew part of $\bgamma$, i.e., 
$\bgamma^A=\left(\bgamma-\bgamma^T\right)\!/2$ and where we used 
the equality $\widehat{\vec{\boldsymbol{\gamma}}}
=-2\boldsymbol{\gamma}^A$. The latter can be shown by noting that 
we have the equalities $\operatorname{Tr}(\widehat{\vec{\boldsymbol{\gamma}}}\hat {\mathbf{u}})=-2\vec{\boldsymbol{\gamma}}\cdot\mathbf{u}=-2\operatorname{Tr}(\boldsymbol{\gamma}\widehat{\mathbf{u}})$ 
for all $\mathbf{u}\in\mathbb{R}^3$.

\paragraph{Bend.} For all $ \mathbf{u} \in \mathbb{R}  ^3 $, we have
\begin{align*}
(\mathbf{n} \times ( \nabla \times \mathbf{n} ) ) \cdot \mathbf{u} 
&=-\nabla _{ \mathbf{n} } \mathbf{n}  \cdot \mathbf{u} 
= -(\mathbf{n} \times \boldsymbol{\gamma}( \mathbf{n} )) \cdot \mathbf{u} 
=-(\mathbf{u} \times \mathbf{n} ) \cdot 
\boldsymbol{\gamma}( \mathbf{n} )\\
&= -\widehat{\mathbf{u}} \mathbf{n}  \cdot 
\boldsymbol{\gamma}( \mathbf{n} ) 
= -\operatorname{Tr}((\widehat{ \mathbf{u}}\mathbf{n})^T 
\boldsymbol{\gamma}\mathbf{n}  )\\
&=\operatorname{Tr} \left((\mathbf{n} \otimes \mathbf{n})
\widehat{ \mathbf{u} } \boldsymbol{\gamma}\right) 
=\operatorname{Tr}\left((\mathbf{I} - {j}/J)
\widehat{ \mathbf{u} } \boldsymbol{\gamma} \right)\\
&=  \operatorname{Tr}( \mathbf{u} \times \boldsymbol{\gamma} )-\frac{1}{J}\operatorname{Tr} (\mathbf{u} \times ( \boldsymbol{ \gamma } {j}))\\
&=\vec{\boldsymbol{\gamma }} \cdot \mathbf{u} - \frac{1}{J} \overrightarrow{\boldsymbol{\gamma }{j}\,} \cdot \mathbf{u},
\end{align*} 
so we get
\[
\mathbf{n} \times ( \nabla \times \mathbf{n} )=\vec{\boldsymbol{\gamma }}- \frac{1}{J}\overrightarrow{\boldsymbol{\gamma }{j}\,} 
\]
and therefore
\[
\|\mathbf{n} \times ( \nabla \times \mathbf{n} )\|^2=\left\|\frac{1}{J}\overrightarrow{\boldsymbol{\gamma }{j}} -\vec{\boldsymbol{\gamma }}\right\|^2=-2\operatorname{Tr}\left(\left(\frac{1}{J}(\bgamma j)^A-\bgamma^A\right)^{\!2}\right).
\]

Summing all the terms, we obtain that the Frank free energy is indeed given by the expression for $ \Psi $ provided in \eqref{FrankMicropolar}.
}

\bigskip

{\footnotesize

\bibliographystyle{new}
%\addcontentsline{toc}{section}{References}

}

\end{document}